\documentclass[aip, reprint]{revtex4-2}

\usepackage{amsmath, amsthm, bm}
\usepackage{amssymb}
\usepackage{mathrsfs}
\usepackage{amsfonts}
\usepackage{graphicx}
\usepackage[caption=false]{subfig}
\usepackage[dvipsnames]{xcolor}
\usepackage{siunitx}
\usepackage{bbold}
\usepackage[version=3]{mhchem}
\usepackage[acronym]{glossaries}
\usepackage{array}
\usepackage{blkarray}

\usepackage[section]{placeins}
\usepackage{indentfirst}

\usepackage[version=3]{mhchem}

\usepackage{xcolor}

\newcommand{\boxthistle}[1]{\colorbox{Thistle}{#1}}
\newcommand{\boxyorange}[1]{\colorbox{YellowOrange}{#1}}
\newcommand{\boxygreen}[1]{\colorbox{YellowGreen}{#1}}

\DeclareSIUnit\au{\text {au}}
\DeclareSIUnit\angstrom{\text {Å}}

\def\mgh{\ce{MgH+}}

\def\eV{\unit{\eV}}

\newacronym{bo}{BOA}{Born-Oppenheimer approximation}
\newacronym{cbo}{CBOA}{cavity Born-Oppenheimer approximation}
\newacronym{dse}{DSE}{dipole self-energy}
\newacronym{pes}{PES}{potential energy surface}
\newacronym{cpes}{cPES}{cavity potential energy surface}
\newacronym{vsc}{VSC}{vibrational-strong coupling}
\newacronym{esc}{ESC}{electronic-strong coupling}
\newacronym{lp}{LP}{lower polariton}
\newacronym{up}{UP}{upper polariton}
\newacronym{jc}{JC}{Jaynes-Cummings}
\newacronym{tc}{TC}{Tavis-Cummings}
\newacronym{fc}{FC}{Franck-Condon}
\newacronym{etc}{ETC}{extended Tavis-Cummings}
\newacronym{rwa}{RWA}{rotating wave approximation}
\newacronym{tls}{TLS}{two-level system}
\newacronym{cs}{CS}{coherent state}
\newacronym{cas}{CASSCF}{complete active space self-consistent field}
\newacronym{mrci}{MRCI}{multiconfiguration reference configuration interaction}
\newcommand{\br}[1]{\langle #1 \vert}
\newcommand{\ke}[1]{\vert #1 \rangle}

\newcommand{\ev}[1]{\langle #1 \rangle}

\begin{document}

\title{Extending the Tavis-Cummings model for molecular ensembles -- Exploring the effects of dipole self energies and static dipole moments}
\author{Lucas Borges}
\affiliation{Department of Physics, Stockholm University, AlbaNova University Center, SE-106 91 Stockholm, Sweden}

\author{Thomas Schnappinger}
\email{thomas.schnappinger@fysik.su.se}
\affiliation{Department of Physics, Stockholm University, AlbaNova University Center, SE-106 91 Stockholm, Sweden}

\author{Markus Kowalewski}
\email{markus.kowalewski@fysik.su.se}
\affiliation{Department of Physics, Stockholm University, AlbaNova University Center, SE-106 91 Stockholm, Sweden}

\date{\today}%

\begin{abstract}
Strong coupling of organic molecules to the vacuum field of a nanoscale cavity can be used to modify their chemical and physical properties. 
We extend the Tavis-Cummings model for molecular ensembles and
show that the often neglected interaction terms arising from the static dipole moment and the dipole self-energy are essential for a correct description of the light-matter interaction in polaritonic chemistry. 
On the basis of a full quantum description, we simulate the excited-state dynamics and spectroscopy of \mgh molecules resonantly coupled to an optical cavity. 
We show that the inclusion of static dipole moments and the dipole self-energy is necessary to obtain a consistent model. 
We construct an efficient two-level system approach that reproduces the main features of the real molecular system and may be used to simulate larger molecular ensembles.
\end{abstract}

\maketitle
\section{Introduction}

Polaritonic chemistry, exploring chemical reactions strongly coupled to a confined electromagnetic field, is an emerging field of research at the interface between quantum optics, quantum chemistry, and materials science~\cite{Ebbesen2016-jx,Garcia-Vidal2021-qe,Ebbesen2023-fd}. By coupling molecules to confined light modes in an optical cavity, the interplay of local excitations and collective excitations in ensembles of quantum emitters gives rise to hybrid states of light-matter known as polaritons~\cite{Flick2017-oc,Feist2018-wg,Ribeiro2018-xd,Herrera2020-bg}. Depending on whether the quantized cavity modes are coupled via their characteristic frequencies to electronic or vibrational degrees of freedom of molecules, the situation is described as \gls{esc} or \gls{vsc}, respectively. Under \gls{esc}, it becomes possible to modify the photochemistry/photophysics of molecules, including photoinduced reactions and electronic spectroscopy~\cite{Schachenmayer2015-hr,Hagenmuller2017-xy,Munkhbat2018-rh,Xiang2018-vn,Groenhof2019-nz,Wellnitz2021-xo,Mony2021-bs,Hutchison2012-od,Mewes2020-qv,Fabri2024-om}. 

The observed effects of molecular \gls{esc} and \gls{vsc} are often discussed phenomenologically by adapting models such as the Rabi model~\cite{Rabi1937-jq}, the Dicke model~\cite{Dicke1954-sq}, the \gls{jc} model~\cite{Jaynes1963-re}, or the \gls{tc} model~\cite{Tavis1967-op}. 
However, all of these models were developed to describe single atoms or atomic ensembles represented by \gls{tls}. 
The addition of nuclear degrees of freedom to the \gls{jc} model and the \gls{tc} model makes it possible to describe processes in the presence of \gls{esc} as non-adiabatic processes~\cite{Kowalewski16jcp,Galego15prx,Fabri2024-om}. These models are also used to simulate large molecular ensemble sizes, due to their simplified description of the coupling and molecules~\cite{Groenhof2019-nz,Hernandez2019-tu,Reitz2020-hl,Schutz2020-en,Tichauer2021-mk,Campos-Gonzalez-Angulo2023-bk}.
Moreover, it has been demonstrated that the concept of non-adiabatic transitions can even be applied to \gls{vsc} \cite{Szidarovszky2021,Fischer2022}.

However, most of these models do not take into account static dipole moments or the feedback of
the light field on the electronic structure.
It has been demonstrated that both static dipoles and a self-consistent treatment of the
electronic structure in the presence of the photon field can be crucial for the description
of polaritonic chemistry \cite{Schafer2020-cb,Sidler2022,Sidler2024,Schnappinger2023-hh,Ruggenthaler2023,Schnappinger_2023,Horak2024-cd}. 
In addition, the \gls{dse} gives rise to a cavity-induced interaction between molecules in an ensemble and depends on the relative molecular orientation in the ensemble \cite{Sidler2024,Schnappinger2023-hh,Haugland2023-rt,Horak2024-cd}. 
In recent years, established electronic structure methods have been generalized to include the effects of quantum light-matter interactions~\cite{Haugland2020-xh,McTague2022-xl,Schnappinger_2023} and used to determine the polaritonic states of molecule-cavity hybrid systems based on the full non-relativistic Pauli-Fierz Hamiltonian~\cite{tokatly2013time,jestadt2019light,Schafer2020-cb,Ruggenthaler2023}. These ab initio methods are more accurate but because of their computational cost, they are limited to single molecules and small molecular ensembles. 

In this manuscript, we build on the framework of the molecular \gls{tc} model
to include both static dipole moments and \gls{dse} contributions, while using 
only field-free molecular properties. 
The starting point is the non-relativistic Pauli-Fierz Hamiltonian in the length gauge. However, since we want to study systems with static dipole moments, the separation into "matter" and "photon" degrees of freedom is no longer trivial~\cite{Castagnola2023-tb,Welakuh2023-ap}. We discuss how this ambiguity of light and matter can be partially circumvented by a \gls{cs} transformation~\cite{Klauder-1985,Haugland2020-xh,Foley-2023,Castagnola2023-tb}. Based on this \gls{cs} Hamiltonian we derive a generalized \gls{tc} Hamiltonian for a molecular system coupled to a single-cavity mode under \gls{esc} conditions. 

As a first test case for the generalized molecular \gls{tc} Hamiltonian, we simulate \mgh molecules resonantly coupled to an optical cavity. 
We investigate the influence of static dipole moments and the influence of the \gls{dse} on the dynamics, and compare the results with the standard \gls{tc} Hamiltonian. Moreover, we analyze the effect of the \gls{cs} transformation that becomes necessary when the molecular ensemble has a nonzero total dipole moment.
In the second step, we calculate and discuss the polaritonic absorption spectra of coupled \mgh-cavity systems. 
On the basis of these results, we construct an effective \glsfirst{tls} model for larger ensembles of \mgh molecules. After optimizing the \gls{tls} parameters, we analyze the structure of this reduced Hamiltonian and study the collective effects induced by the interaction with the cavity mode.
 
\section{Theory and models}

In the following, we make use of the non-relativistic Pauli-Fierz Hamiltonian in the length gauge representation~\cite{tokatly2013time,jestadt2019light,Schafer2020-cb,Ruggenthaler2023} to describe the interaction of molecules with the confined electromagnetic field. 
Atomic units ($\hbar=4\pi\varepsilon_0=m_e=1$) are used throughout the paper unless otherwise noted, and bold symbols denote vectors.

The corresponding Pauli-Fierz Hamiltonian $\hat{H}_{\text{PF}}$ for a single cavity mode within the \gls{bo} takes the form
\begin{align}\label{eq:HPF}
\hat{H}_{\text{PF}} = &\hat{T}_{nuc} + \hat H_{el} + \omega_c\left(\hat{a}^\dagger\hat{a} + \frac{1}{2}\right) \\
&- \sqrt{\frac{\omega_c}{2}}(\hat{a}^\dagger + \hat{a})\left(\bm{\lambda} \cdot \hat{\bm{\mu}}\right) + \frac{1}{2} \left(\bm{\lambda} \cdot \hat{\bm{\mu}}\right)^2, \nonumber
\end{align}
where $\hat{T}_{nuc}$ is the nuclear kinetic energy operator and $\hat H_{el}$ is the electronic Hamiltonian, both defining the Hamiltonian of the molecular subsystem $\hat{H}_{m}$. 
The third term in Eq.~\eqref{eq:HPF} is purely photonic and describes the single-cavity mode as a quantum harmonic oscillator with frequency $\omega_c$. 
The operators $\hat{a}^\dagger$ and $\hat{a}$ are the bosonic photon creation and annihilation operators~\cite{Schleich2001}. 
The fourth term describes the dipole coupling of the photon mode and  molecular
degrees of freedom, where $\bm{\hat{\mu}}$ is the molecular dipole moment operator and 
\begin{equation}
 \label{eq:lam}
\bm{\lambda} =  \bm{e} \lambda =  \bm{e}  \sqrt{\frac{4 \pi}{V_{c}}}\,,
\end{equation}
is the coupling parameter.
Here, $V_c$ is the cavity quantization volume and $\bm e$ is the polarization vector of
the photon mode.
The last term in Eq.~\eqref{eq:HPF} is the \gls{dse} contribution~\cite{Rokaj_2018,Schafer2020-cb,Sidler2022}, which describes the self-polarization of the molecule-cavity system. 

To compare the dynamics of different ensembles with a varying number of molecules,
we keep the collective coupling strength $\lambda_c$ constant by scaling the single molecule coupling strength $\bm{\lambda}$ with $1/\sqrt{N_{mol}}$.
\begin{equation}
\label{eq:coupling}
\bm{\lambda} = \frac{\lambda_c}{\sqrt{N_{mol}}} \bm{e}\,,
\end{equation}
where $N_{mol}$ is the number of molecules and $\lambda_c$ is then treated as a tunable coupling parameter.

\subsection{The coherent state transformation}
\begin{figure}
     \centering
\includegraphics[width=8.0cm]{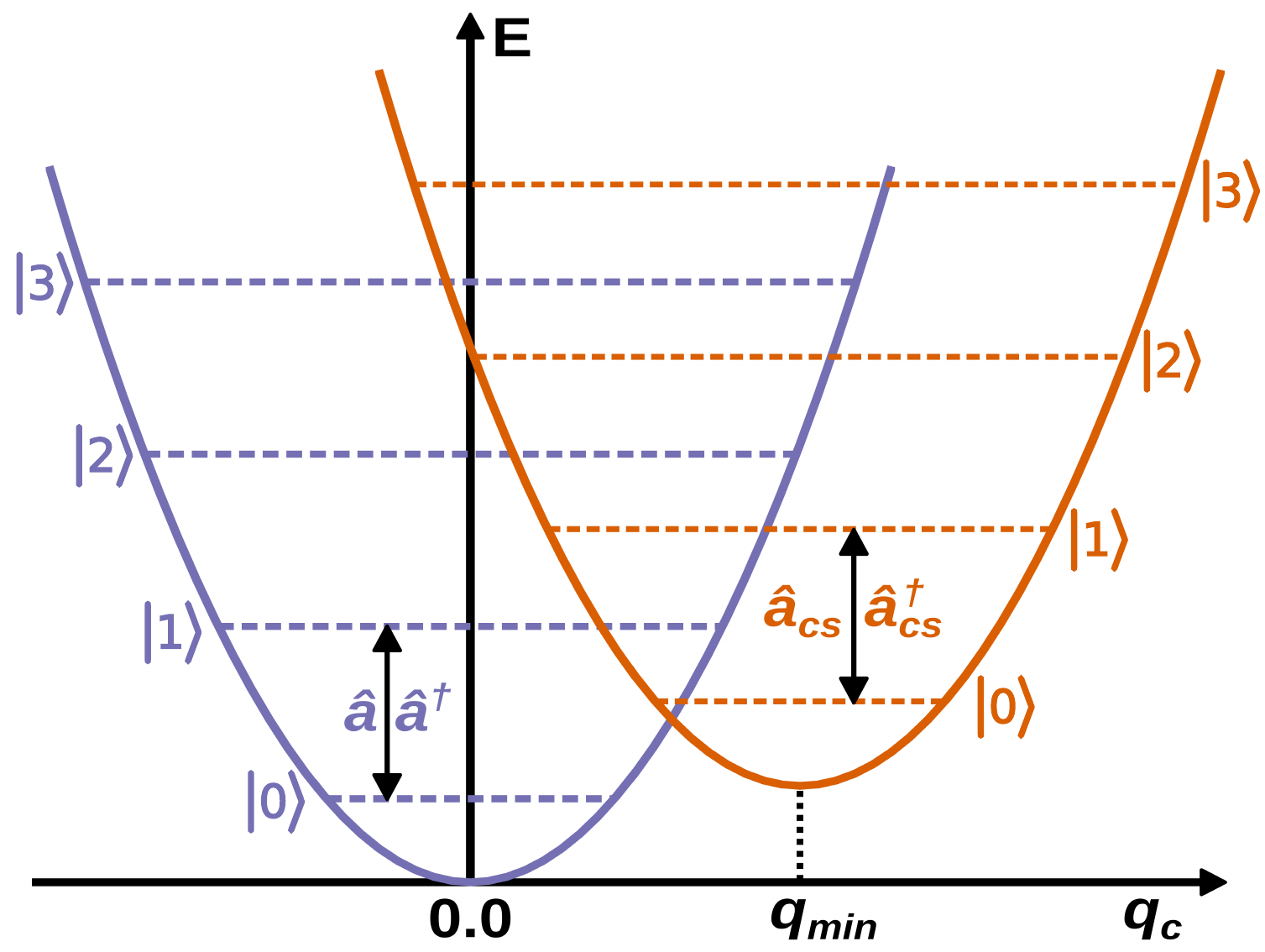} 
    \caption{Schematic representation of the harmonic potential for an uncoupled cavity mode (purple) and a cavity mode coupled to a molecular system with a static dipole moment (orange). Due to the light-matter interaction, the potential is shifted in energy and the displacement field coordinate $q_c$. Photonic eigenstates are indicated by colored dashed lines, and $\ke{0}\rightarrow\ke{1}$ transitions are marked with the corresponding creation and annihilation operators.} 
\label{fig:disp_cs_basis}
\end{figure}
The transformation of the Pauli-Fierz Hamiltonian to the dipole gauge leads to a mixing
of the light and matter degrees of freedom~\cite{Castagnola2023-tb,Welakuh2023-ap}.
The consequence is a shift of the photon mode, which arises for molecular ensembles with a static dipole moment.
In Eq.~\eqref{eq:HPF} the photonic part and the light-matter interaction are written in terms of the photonic creation and annihilation operators $\hat{a}^\dagger$ and $\hat{a}$, which are defined for an empty cavity mode. 
To visualize how $\hat{a}$ changes for a system with a static dipole moment,
we express the photon mode in terms of photon displacement coordinates $q_c$ and $p_c$
~\cite{Schleich2001,Kowalewski2016-zo}: 
\begin{equation} 
q_c = \frac{1}{\sqrt{2\omega_c}}\left(\hat a + \hat a^\dagger\right), \quad
p_c = -i\sqrt{\frac{\omega_c}{2}}\left(\hat a - \hat a^\dagger\right)\,.
\end{equation} 

The photon mode potentials for a coupled (orange) and an uncoupled (purple) case are shown in Fig.~\ref{fig:disp_cs_basis}.
For the uncoupled cavity mode (purple in Fig.~\ref{fig:disp_cs_basis}) the minimum of the harmonic potential is at $q_c = 0$. The corresponding creation and annihilation operators $\hat{a}^\dagger$ and $\hat{a}$ are the usual ladder operators of the quantum harmonic oscillator.
The coupling of a molecular system with a static dipole moment affects both the molecule and the cavity mode. The light-matter interaction shifts the photon mode potential in $q_c$ (Fig.~\ref{fig:disp_cs_basis}, orange). The minimum of the shifted harmonic potential for a given nuclear configuration $\bm{R}$ is at,
\begin{equation}
q_{min}(\bm{R}) = - \frac{ \bm{\lambda} \cdot  \bigl\langle \bm{\hat{\mu}} \bigr\rangle_{0} }{\omega_c},
\end{equation}
with $\bigl\langle \bm{\hat{\mu}} \bigr\rangle_{0} \equiv  \bigl\langle \bm{\hat{\mu}} \bigr\rangle_{0}(\bm{R})$ being the static dipole moment function of the molecular ground state~\cite{Schnappinger_2023,Angelico2023-qn}. 
It becomes clear that $\hat{a}^\dagger$ and $\hat{a}$ are no longer
valid ladder operators for the shifted cavity field potential. The same holds for the number operator $\hat{N} = \hat{a}^{\dagger} \hat{a}$, which no longer produces valid photon numbers. Note, that the \gls{cs} transformation becomes relevant
as soon as the ensemble exhibits a static dipole moment. 
For a more detailed discussion of this topic, we refer the reader to references~\cite{Flick2017-oc,Sidler2024,Schnappinger_2023,Welakuh2023-ap}. 

To compensate for the shift in the photon mode, the coherent state transformation
is used ~\cite{Klauder-1985,Haugland2020-xh,Foley-2023,Castagnola2023-tb}.
The unitary transformation 
\begin{equation}
\hat{U}_{cs}(\bm{R}) =  e^{ z\left( \hat{a}^\dagger - \hat{a} \right)}  \  \text{ with} \ \ z(\bm{R}) = q_{min} \sqrt{\frac{\omega_c}{2}}  = - \frac{  \bm{\lambda} \cdot  \bigl\langle \bm{\hat{\mu}}\bigr\rangle_{0}}{\sqrt{2\omega_c}}\,.
\end{equation}
yields new annihilation and creation operators $\hat{a}_{cs}^\dagger(\bm{R})$ and $\hat{a}_{cs}(\bm{R})$ which now depend on the nuclear configuration through the
static dipole moments.
These operators can be expressed in terms of the original operators $\hat{a}^\dagger$ and $\hat{a}$:
\begin{equation}
 \label{eq:cs_trans}
 \begin{split}
\hat{a}^\dagger_{cs} (\bm{R}) = \hat{U}_{cs}\hat{a}^\dagger \hat{U}_{cs}^\dagger & = \hat{a}^\dagger - z = \hat{a}^\dagger +  \frac{  \bm{\lambda} \cdot  \bigl\langle \bm{\hat{\mu}}\bigr\rangle_{0}(\bm{R}) }{\sqrt{2\omega_c}}, \\
\hat{a}_{cs}(\bm{R}) = \hat{U}_{cs}\hat{a}\hat{U}_{cs}^\dagger & = \hat{a} - z = \hat{a} +  \frac{  \bm{\lambda} \cdot  \bigl\langle \bm{\hat{\mu}}\bigr\rangle_{0}(\bm{R}) }{\sqrt{2\omega_c}}\,.
 \end{split}
\end{equation}
Applying the same \gls{cs} transformation to the full Pauli-Fierz Hamiltonian $\hat{H}_{\text{PF}}$ yields the corresponding operators in the \gls{cs} basis:
\begin{align}
\label{eq:CSPF}
\hat{H}_{cs} = &\hat{U}_{cs}\hat{H}_\text{PF} \hat{U}_{cs}^\dagger 
 = \hat{H}_{m} + \omega_c \left(\hat{a}^{\dagger} \hat{a} + \frac{1}{2} \right) \\
 &-\sqrt{\frac{\omega_c}{2}}(\hat{a}^\dagger + \hat{a})\left(\bm{\lambda} \cdot \bm{\Tilde{\mu}}\right) + \frac{1}{2} \left(\bm{\lambda} \cdot \bm{\Tilde{\mu}}\right)^2, \nonumber
\end{align}
with $ \bm{\tilde{\mu}} = \hat{\bm{\mu}}-\ev{\hat{\bm{\mu}}}_0$ describing the change in dipole moment with respect to the ground state. 
As a consequence, the \gls{dse} contribution takes the following form:
\begin{align}
\frac{1}{2} \left(\bm{\lambda} \cdot \bm{\Tilde{\mu}}\right)^2  = &
\frac{1}{2}\left(\bm{\lambda} \cdot \hat{\bm{\mu}}\right)^2 
- \left(\bm{\lambda} \cdot \hat{\bm{\mu}}\right)
\left(\bm{\lambda} \cdot \ev{\hat{\bm{\mu}}}_0\right) \\
& + \frac{1}{2}\left(\bm{\lambda} \cdot \ev{\hat{\bm{\mu}}}_0\right)^2. \nonumber
\end{align}
In the same way, $\hat{N}$ can also be transformed into the \gls{cs} basis.

In the following, we will use the Pauli-Fierz Hamiltonian in the \gls{cs} basis $\hat{H}_{cs}$ to describe the cavity-molecule systems, unless otherwise noted. Assuming that the dipole moment is oriented in parallel to the polarization axis of the cavity mode, the scalar product $\bm{\lambda}\cdot\tilde{\bm{\mu}}$ is reduced to the simple product $\lambda\tilde{\mu}$. 
The Hamiltonians shown in Eq.~\eqref{eq:HPF} and in Eq.~\eqref{eq:CSPF} are formally equivalent in the complete basis limit~\cite{Foley-2023,Vu_2024}. 
However, the photonic states described by $\hat{a}^\dagger$ and $\hat{a}$ are not necessarily proper annihilation or creation operators of the coupled cavity-molecule system.
Note that we use Eq.~\eqref{eq:CSPF} with the \gls{bo} applied throughout the rest of the paper. All operators are then operators that act on the electronic eigenstates, the nuclear coordinates, and the Fock-states of the photon field.

\subsection{The extended Tavis-Cummings Hamiltonian}
\begin{figure}
\centering
\includegraphics[width=8.0cm]{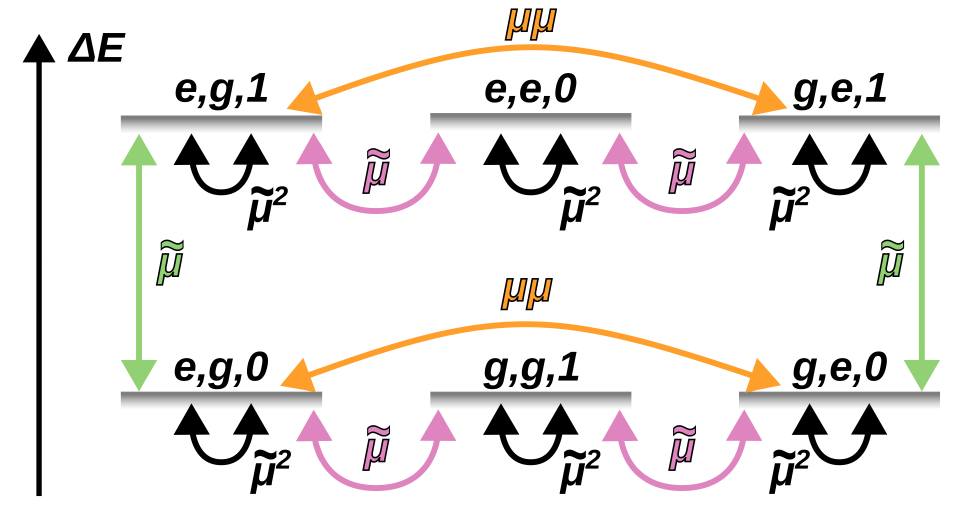}
\caption{Scheme for the couplings between the bare-states of a coupled system of two molecules and a cavity mode, where $\tilde\mu$ stands for the static dipole coupling, $\tilde\mu^2$ the \gls{dse} shifts and $\mu\mu$ the intermolecular excited states couplings arising from the \gls{dse}.}
\label{fig:diagram_v2mol}
\end{figure}

In this manuscript we do not solve the coupled electron-polariton part of $\hat{H}_{\text{CS}}$ self-consistently, but use field-free molecular properties, such as \glspl{pes} and dipole moment functions in an adiabatic basis assuming the \gls{bo}, in the derivation of the generalized Tavis-Cummings Hamiltonian. 
The molecular \gls{tc} model was formulated under the assumption that the interacting molecules do not have a static dipole moment. In the following, we extend the molecular \gls{tc} model to include static dipole moments and the \gls{dse} terms of Eq.~\ref{eq:HPF}. We will refer to this generalized model as the molecular \gls{etc} ansatz.

The molecular Hamiltonian $\hat{H}_m^{(i)}$ of the $i$th molecule has two electronic states, a ground state $g$ and the first excited state $e$.
\begin{equation}
\hat{H}_m^{(i)} = \hat{T}_{nuc} + V_g(\bm{R}_i)\hat{\sigma}^{(i)}\hat{\sigma}^{(i)\dagger} + V_e(\bm{R}_i)\hat{\sigma}^{(i)\dagger}\hat{\sigma}^{(i)}\,, \label{eq:Hm_1mol}
\end{equation}
where $\bm{R}_i$ is a set of nuclear coordinates of the $i$th molecule.
The operators $\hat{\sigma}^{(i)} = \ke{g_i}\br{e_i}$ and $\hat{\sigma}^{(i)^\dagger} = \ke{e_i}\br{g_i}$ annihilate and create, respectively, an excitation in the electronic subspace on the $i$th molecule, defined by the ground and first excited state \glspl{pes} $ V_g(\bm{R})$ and $V_e(\bm{R})$, respectively. 
The corresponding dipole moment and squared dipole moment operators for the individual molecule of the $ith$ molecule can be expressed as follows:
\begin{align}
\hat{\mu}^{(i)} = &
    \mu_{gg}\hat{\sigma}^{(i)}\hat{\sigma}^{(i)\dagger} + 
    \mu_{ee}\hat{\sigma}^{(i)\dagger}\hat{\sigma}^{(i)}\\
    &+ \mu_{eg}\left( \hat{\sigma}^{(i)} + \hat{\sigma}^{(i)\dagger} \right)\,,    \nonumber
\end{align}
\begin{align}
\left(\hat{\mu}^{(i)}\right)^2 = & 
  \mu^2_{gg}\hat{\sigma}^{(i)}\hat{\sigma}^{(i)\dagger}  + \mu^2_{ee}\hat{\sigma}^{(i)\dagger}\hat{\sigma}^{(i)} \\
  &+\mu^2_{eg}\left( \hat{\sigma}^{(i)} + \hat{\sigma}^{(i)\dagger} \right)\,,    \nonumber    
\end{align}
where $\mu_{mn}\equiv\ev{\mu}_{mn}(\bm{R_i})$ and $\mu^2_{mn}\equiv\ev{\mu^2}_{mn}(\bm{R_i})$ are the $\bm{R_i}$ dependent dipole matrix elements, and squared dipole moments between electronic states $m$ and $n$ respectively.
The total dipole moment operator $\tilde{\mu}$ after the \gls{cs} transformation reads:
\begin{equation}
\tilde{\mu} = \sum_{i=1}^{N} \hat{\mu}^{(i)} - \ev{\hat{\mu}}_0,
\end{equation}
where $\ev{\hat{\mu}}_0$ is the ground state static dipole moment of the whole ensemble. The corresponding squared dipole operator is given by
\begin{equation}
\label{eq:mu2_2m}
   \tilde{\mu}^2 = \sum_{i=1}^{N} \left(\hat{\mu}^{(i)}\right)^2 
    - 2\hat{\mu}^{(i)}\ev{\hat{\mu}}_0  + \sum_{j \neq i}^{N}  \hat{\mu}^{(i)}\hat{\mu}^{(j)} + \ev{\hat{\mu}}_0^2.
\end{equation}
The first two terms are operators acting locally on each molecule. On the contrary, the third term of Eq.~\eqref{eq:mu2_2m} describes an intermolecular interaction by directly connecting the dipole moment operators $\hat{\mu}^{(i)}$ and $\hat{\mu}^{(j)}$ of two molecules. This interaction of two molecules induced by \gls{dse} has been shown to play an important role in the description of molecular ensembles under \gls{vsc}~\cite{Sidler2024,Schnappinger_2023}.

The total wave function of the coupled ensemble is represented as a tensor product of the wave function of each molecule and the Fock states of the photon mode.
Here, we truncate the wave function to a maximum of two excitations. 
Each molecule is by definition limited to a maximum of one excitation.
The resulting product wave function for $N$ molecules reads:
\begin{align}\label{eq:bare_states}
\ke{\Psi;n_p} : & \left\{ \ke{G;0},\ke{G;1}, \ke{E^{(i)};0},  \right.\\
&\left. \dots, \ke{G;2},  \ke{E^{(i)};1},   \dots,  \ke{\mathscr{E}^{(i,j)};0}, \dots \right\} \nonumber
\end{align}
where $\ke{G} \equiv \ke{g_1, \dots }$ is the collective molecular ground state. 
The $2N$ states of the form $\ke{E^{(i)};n} \equiv \ke{g_1,g_2, e_i, \dots;n}$ are described by a single excited molecule $i$ and $n$ photons, and $N(N-1)/2$ additional states of the form $\ke{\mathscr{E}^{(i,j)};0} \equiv \ke{g_1,e_i,e_j,\dots;0}$ are characterized by two excited molecules.

In Eq.~\eqref{H_Nmol_matrix}, we show the schematic structure of the matrix representing the light-matter interaction terms of $\hat{H}_{cs}$, 
which consist of linear dipole coupling and the \gls{dse} terms.
Since the matrix is symmetric, only the upper triangle is shown, and prefactors are excluded for improved clarity. Since we are interested in the dynamics in the first excitation manifold, we do not show all coupling terms within the $\ke{\mathscr{E}^{(i,j)};0}$ states.
\begin{widetext}
\begin{equation} \label{H_Nmol_matrix}
\begin{blockarray}{cccccccccccc} 
 & \ke{G;0} & \ke{G;1} & \ke{E^{(1)};0} &\cdots & \ke{E^{(N)};0} & \ke{G;2} & \ke{E^{(1)};1} &\cdots & \ke{E^{(N)};1} & \ke{\mathscr{E}^{(i,j)};0}\\
\begin{block}{@{} r @{\hspace{1ex}}  @{\quad} (ccccccccccc) }
  \br{G;0} & \left(\lambda\tilde{\mu}\right)^2_{G} & 0 & 0 & \cdots & 0 & 0 & 0 & \cdots & 0 & 0\\
 \br{G;1} &  & \left(\lambda\tilde{\mu}\right)^2_{G}& \boxthistle{$\lambda\tilde{\mu}_{eg}$} & \cdots &  \boxthistle{$\lambda\tilde{\mu}_{eg}$} & 0 & 0 & \cdots & 0 & 0\\ 
 \br{E^{(1)};0} &  &  & \left(\lambda\tilde{\mu}\right)^2_{E}& & \boxyorange{$\left(\lambda\tilde{\mu}\right)^2_{ee}$} & 0 & \boxygreen{$\lambda\tilde{\mu}_{ee}$} & & 0 & 0\\
 \vdots~\quad\quad &&&&\ddots&&\vdots&&\ddots&&\vdots \\
 \br{E^{(N)};0} & &  &  &  & \left(\lambda\tilde{\mu}\right)^2_{E}  & 0  & 0 & & \boxygreen{$\lambda\tilde{\mu}_{ee}$} & 0\\
  \br{G;2} &  &  & &  & &  \left(\lambda\tilde{\mu}\right)^2_{G} &  \boxthistle{$\lambda\tilde{\mu}_{eg}$} & \cdots & \boxthistle{$\lambda\tilde{\mu}_{eg}$} &  0 \\
  \br{E^{(1)};1} &  & &  &  & & & \left(\lambda\tilde{\mu}\right)^2_{E}& &\boxyorange{$\left(\lambda\tilde{\mu}\right)^2_{ee}$} & \boxthistle{$\lambda\tilde{\mu}_{eg}$} \\
  \vdots~\quad\quad &&&&&&&&\ddots&&\vdots\\
 \br{E^{(N)};1} &  &  & &  && & &  & \left(\lambda\tilde{\mu}\right)^2_{E} & \boxthistle{$\lambda\tilde{\mu}_{eg}$}\\
 \br{\mathscr{E}^{(i,j)};0} &  & & &  &  & & &  & & \left(\lambda\tilde{\mu}\right)^2_{\mathscr{E}} \\
\end{block}
\end{blockarray}
\end{equation}
\end{widetext}

To further reduce complexity, the \gls{rwa} has been applied, which removes all rapidly oscillating terms~\cite{QOiPS_20001}. For validation, we performed benchmark calculations with and without the \gls{rwa}, the results are shown in Section S1 of the Supporting Information. 

The linear dipole interactions create off-diagonal terms that can be categorized into two groups: the first group (highlighted in purple in Eq.~\eqref{H_Nmol_matrix}) corresponds to the conventional \gls{tc} coupling terms that couples different electronic states. 
The second group of linear couplings (highlighted in green in Eq.~\eqref{H_Nmol_matrix}), which are not part of the standard \gls{tc} Hamiltonian,
couples different vibrational states within the same electronic state.
This coupling term is zero for all states formed by the ensemble ground state due to the \gls{cs} transformation. Note that vibrational states in the electronic ground
state are coupled indirectly through the dependence of $\hat a_{cs}$ on $\bm{R}$.
The \gls{dse} terms yield two different types of terms.
The first type are the diagonal elements $\left(\lambda\tilde{\mu}\right)^2_{G}$, $\left(\lambda\tilde{\mu}\right)^2_{E}$ and $\left(\lambda\tilde{\mu}\right)^2_{\mathscr{E}}$ that lead to a state-specific energy shift.
The second group of \gls{dse} contributions (marked orange) connects states with the same photon number but with electronic excitations located on different molecules. 
These terms are a direct consequence of the intermolecular dipole-dipole interaction
of Eq.~\eqref{eq:mu2_2m}. 
The corresponding matrix elements have the following form:  
\begin{equation} \label{eq:intermolecularDSE}
    \left(\lambda\tilde{\mu}\right)^2_{ee} = \lambda^2 \mu^{(i)}_{eg}(\bm{R}_i) \mu^{(j)}_{ge}(\bm{R}_j)\,,
\end{equation}
and show that the molecular excitations can be exchanged through the cavity mode by means of the \gls{dse}.
A detailed derivation of all interaction terms for the case of $N$ molecules can be found in Section S1 of the Supporting Information. Note that increasing the number of molecules in this model
increases the size of the matrix $\hat{H}_{cs}$, but does not introduce new types of interaction. 

All relevant interactions are depicted schematically for the single excitation manifold in Fig.~\ref{fig:diagram_v2mol}.
All states within the first excitation manifold
are directly coupled by either the linear dipole interaction (pink)
or the intermolecular \gls{dse} contribution (orange).

\begin{figure}
\centering
\includegraphics[width=8.5cm]{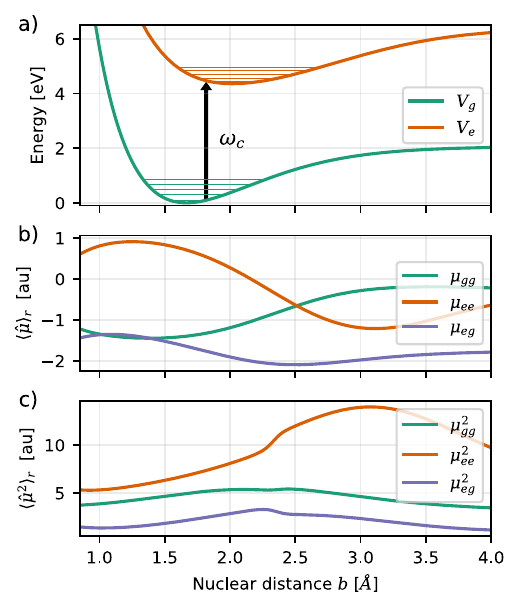}
\caption{a) Bare ground ($\Sigma_0 \equiv g$) and excited ($\Sigma_1 \equiv e$) electronic potential energy surfaces of \mgh. b) Static and transition dipole moment functions ($\ev{\mu}_{gg}$, $\ev{\mu}_{ee}$ and $\ev{\mu}_{ge}$) along the molecular bond ($z$ axis) and c) corresponding squared dipole moment functions ($\ev{\mu^2}_{gg}$, $\ev{\mu^2}_{ee}$ and $\ev{\mu^2}_{eg}$).}
\label{fig:MoleculeDetails}
\end{figure}

\section{Computational details}

All electronic structure calculations of \mgh are performed with the MOLPRO program package~\cite{MOLPRO-WIREs} version 2021.2 \cite{Knowles1988-qu,Knowles1985-nr,Werner1985-jk,Werner1988-sk} at the CAS(6/9)/MRCI/aug-cc-pVQZ \cite{Kendall1992-wu} level of theory with six active electrons in nine orbitals \cite{Davidsson2020-qb,Davidsson2020-bs,Kahra12np}.
In total, five electronic states are included in the state-average procedure. 
The static and transition dipole moments are obtained directly from MOLPRO, while the squared dipole moments are calculated using a resolution of identity approach~\cite{Gudem2021-um,Couto2022-uv,Weight2023-ma}:
\begin{equation}
  \mu^2_{ij} (\bm{R}) \approx  \sum_{k=1}^5    \mu_{ik} (\bm{R}) \mu_{kj} (\bm{R})
\end{equation}
Here $i$, $j$, and $k$ refer to electronic states, and the sum 
runs over the all five states involved in the state averaging procedure (see Section S5 of the Supporting Information for details on the convergence of the squared dipole moments).

All necessary properties (see Fig.~\ref{fig:MoleculeDetails}), 
such as \glspl{pes} and dipole moments, are
calculated on a coarse grid between $R=0.8$\,\AA\ and $R=4.0$\,\AA\
and interpolated to a finer grid.
Both states exhibit a static dipole moment and a transition dipole moment along the molecular bond, see Fig.~\ref{fig:MoleculeDetails}(b). 
The corresponding squared dipole moments calculated using the resolution of identity approach are shown in Fig.~\ref{fig:MoleculeDetails}(c). 
\begin{table}
\caption{\label{tbl:grid_data} Details of the grid and simulation parameters. The number of points $N$ is given for each dimension of the grid. The minimum and maximum values of the internuclear distance $R$, the propagation time, and the time step are given in atomic units. }
  	\centering
  \begin{tabular}{l c c c c c c}
   \hline \hline
& $N$  & $R_{min}$ [au] & $R_{max}$ [au] & $\Delta t$ [au] & $t_{max}$ [fs] \\
\hline
\ce{(MgH+)1} & 128  & 1.61 & 7.56 & 5.0 &  500  \\

\ce{(MgH+)2} & 64x64 & 1.61 & 6.61 & 5.0 &  500  \\

\ce{(MgH+)3} & 64x64x64 & 1.61 & 6.61 &  5.0 &  500  \\
\hline \hline
 \end{tabular}
\end{table}
The two- and three-dimensional surfaces for the ensemble of two and three \mgh molecules are constructed from the molecular \glspl{pes}. Details of all three grids can be found in TABLE~\ref{tbl:grid_data}. 

The cavity frequency $\omega_c = \SI{4.322}{\electronvolt}$ is chosen to be resonant to the energy difference between the first vibrational states of each potential ($\ke{g,v=0}\rightarrow\ke{e,v=0}$ transition), which is indicated by the arrow in Fig.~\ref{fig:MoleculeDetails}(a).
In addition to the standard \gls{tc} coupling schema and our \gls{etc} Hamiltonian we extend the molecular \gls{tc} model only with static dipole moments or \gls{dse} contributions. 
The latter two are only used for benchmarking purposes. 

\begin{figure*}
\includegraphics[width=0.95\textwidth]{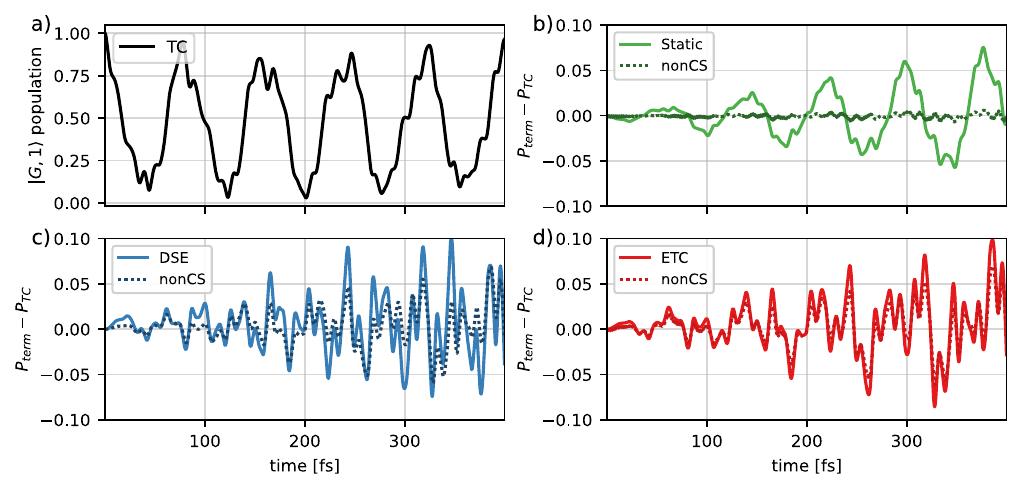} 
\caption{Time evolution of the $\ke{G;1}$ population for a single \mgh molecule coupled to the cavity mode for a) the molecular \gls{tc} model, b) populations differences to the molecular \gls{tc} model by the inclusion of only the static dipole coupling interactions terms, c) only the \gls{dse} terms, and d) the molecular \gls{etc} model considering both terms. The corresponding results without the \gls{cs} transformation are indicated by dotted lines. All propagations were performed with cavity resonance frequency $\omega_c = \SI{4.322}{\electronvolt}$ and coupling strength $\lambda_{c} = \SI{6.9e-3}{au}$.}
\label{fig:pop_dyn_1mgh}
\end{figure*}
\begin{figure*}
\centering
\includegraphics[width=0.95\textwidth]{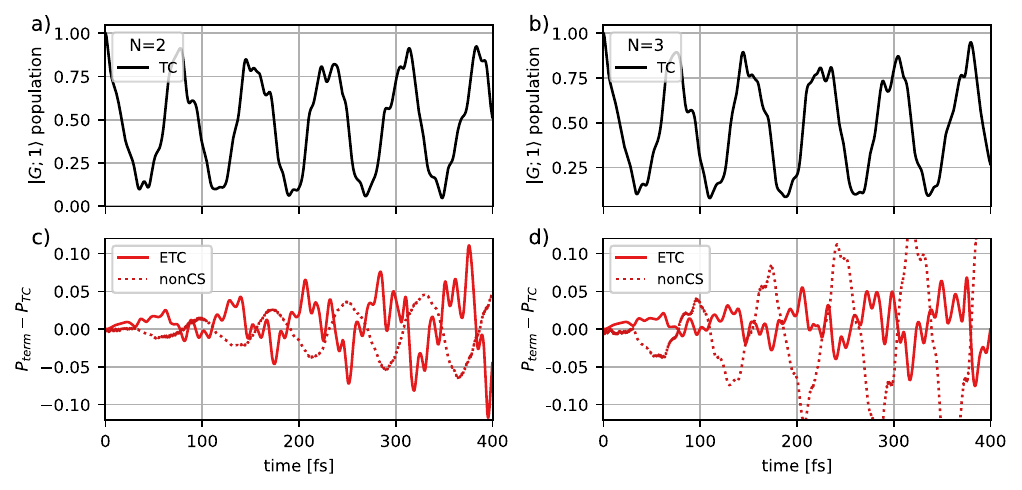} 
\caption{Time evolution of the $\ke{G;1}$ population for two \mgh molecules, a and c), and for three \mgh molecules, b) and d), coupled to the cavity mode. 
Figures a) and b) show the results for molecular \gls{tc}. 
Figures c) and d) show the difference in the $\ke{G;1}$ population between the molecular \gls{tc} and \gls{etc} Hamiltonian, which includes the static dipole coupling interactions and the \gls{dse} terms. 
The corresponding results without the \gls{cs} transformation are shown by dotted lines. All simulations were performed with cavity excitation $\omega_c = \SI{4.322}{\electronvolt}$ and a collective coupling strength $\lambda_{c} = \SI{6.9e-3}{au}$.}
\label{fig:pop_dyn_multi_mgh}
\end{figure*}

To evaluate the influence of the new terms introduced in the molecular \gls{etc} model,
we compare the dynamics of small molecular ensembles to the molecular \gls{tc} model.
For all Hamiltonians, the excited state dynamics has been simulated by numerically propagating the time-dependent Schr\"odinger equation with 
the Arnoldi propagation scheme \cite{Smyth1998-dv}.
The vibrational eigenfunctions of the uncoupled potentials are obtained using the imaginary time propagation method~\cite{Kosloff1986-ow}. 
The optimized ground state wave function $\ke{G,0}$ is used to initiate coupled dynamics in the $\ke{G,1}$ state. 
The grid-based quantum dynamics simulations are performed with the QDng quantum dynamics program package~\cite{Kowalewski2024-ue}. All calculations were performed in a reproducible environment using the Nix package manager together with NixOS-QChem \cite{nix} (commit f803c22214) and Nixpkgs (nixpkgs, 23.05, commit 5550a85a08). 

\section{Population dynamics of different light-matter Hamiltonians} \label{sec:pop_mol}
Let us first analyze the effect of the additional terms in the molecular \gls{etc} model on the dynamics of a single \mgh molecule and compare its results with those of the standard molecular \gls{tc} model. 
These terms are the static dipole moment coupling and the contribution of \gls{dse} as well as the changes induced by the \gls{cs} transformation.
Figure \ref{fig:pop_dyn_1mgh} shows the population dynamics of $\ke{G;1}$ using the standard molecular \gls{tc} model, as well as the changes in the population caused by adding the additional terms mentioned above to the light-matter Hamiltonian and the differences caused by the \gls{cs} transformation.

For the molecular \gls{tc} model, the population dynamics of the $\ke{G;1}$ state is characterized by Rabi oscillations with a period of 80\,fs, which corresponds to an effective Rabi frequency of 52\,meV/$\hbar$ (see Fig.~\ref{fig:pop_dyn_1mgh}~a)). 
The fine structure of the oscillations is caused by the motion of the nuclear wave packet
in the $\ke{G;1}$ and the $\ke{E;0}$ states. 
By including additional coupling terms in the Hamiltonian,
the Rabi oscillations are preserved, but the resulting population dynamics are changed compared to the molecular \gls{etc} simulation, which can be seen in
Figs.\ \ref{fig:pop_dyn_1mgh}~b) and \ref{fig:pop_dyn_1mgh}~c).
The observed differences are mainly caused by a difference in the observed Rabi frequency, which also explains the increase in the difference with increasing simulation time. 
Taking into account only the static dipole interaction terms (Fig.~\ref{fig:pop_dyn_1mgh}~b)) or only the \gls{dse} contributions (Fig.~\ref{fig:pop_dyn_1mgh}~c)),
the frequency difference leads to a maximum population difference of 0.075 and 0.10, respectively, within the initial 400~fs. 
The population difference, when only the static dipole interaction terms are included, shows a slow oscillating pattern with a frequency comparable to the Rabi frequency, whereas for the case of only \gls{dse} contributions much faster oscillations are observed. 
For the molecular \gls{etc} model (Fig.~\ref{fig:pop_dyn_1mgh}~d)) these two contributions are combined resulting in a maximum population difference of 0.10  in the same time window.
Figures \ref{fig:pop_dyn_1mgh}~b) through d) also show the cases where the \gls{cs} transformation is not applied.  
In the case of the molecular \gls{etc} Hamiltonian (Fig.~\ref{fig:pop_dyn_1mgh}~d)) dropping the \gls{cs} transformation results in a negligible error. 
However, if the molecular \gls{etc} Hamiltonian is only extended by static dipole
terms the result is almost identical to the molecular \gls{tc} model, see Fig.~\ref{fig:pop_dyn_1mgh}~b). In contrast, comparing the \gls{dse}-only results (Fig.~\ref{fig:pop_dyn_1mgh}~c) with and without the \gls{cs} transformation gives a comparable result. This behavior, and the negligible error for the molecular \gls{etc} Hamiltonian, indicates that the population difference between the \gls{etc} and \gls{tc} models is mostly determined by the \gls{dse} contribution. 

The population dynamics of the $\ke{G;1}$ state is visualized in Fig.~\ref{fig:pop_dyn_multi_mgh} for the case of two and three \mgh molecules coupled to a cavity to examine how the molecular \gls{etc} Hamiltonian and the \gls{cs} transformation change the results of the standard molecular \gls{tc} model.
The molecules are assumed to be oriented in parallel. 
The population dynamics obtained with the molecular \gls{etc} model and the observed Rabi oscillations are qualitatively similar for one, two, and three \mgh molecules.
Note that the collective coupling strength is kept constant, 
due to the rescaling of the single-particle coupling strength by $1/\sqrt{N_{Mol}}$
(see Eq.~\eqref{eq:coupling}). 
However, the differences between molecular \gls{tc} and \gls{etc} Hamiltonian are affected by the increase in the number of molecules. 
Going from a single molecule (Fig.~\ref{fig:pop_dyn_1mgh}~c) solid line) to two molecules (Fig.~\ref{fig:pop_dyn_multi_mgh}~c) solid line) the deviation between the \gls{tc} model and the \gls{etc} model remains comparable, while for three molecules (Fig.~\ref{fig:pop_dyn_multi_mgh}~d) solid line) the size of the deviation is reduced. 
The corresponding populations difference due to the inclusion of only the \gls{dse} or only the static dipole moment are shown in Fig.~S3 in the Supporting Information. Interestingly, we observe that the influence of the static dipole moment contribution decreases with the number of molecules faster than the \gls{dse}, which can be attributed to the \gls{cs} transformation. 
If the transformation is not performed (see Figs.~\ref{fig:pop_dyn_multi_mgh}(c) and (d), dotted lines) the difference between the molecular \gls{tc} and \gls{etc} model becomes larger for an
increasing number of molecules, which can be explained by the increasing total
dipole moment of the ensemble.
In summary, in the case of a few molecules, the non-trivial interplay of \gls{dse} contributions and the presence of static dipole moments define a situation where none of the terms can be simply neglected. 
In particular, intermolecular dipole-dipole interactions due to \gls{dse} in the molecular \gls{etc} Hamiltonian (Eq.~\eqref{eq:intermolecularDSE}) play an important role, see section S1 in the Supporting Information.

\section{Polaritonic absorption spectra}\label{sec:spec_mol}

\begin{figure}
\centering
\includegraphics[width=8.5cm]{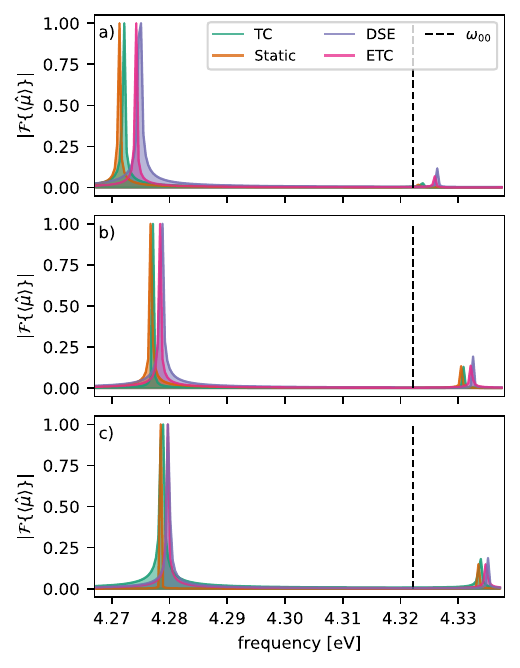}
\caption{Relevant part of the polaritonic absorption spectra for a) 1, b) 2, and c) 3 \mgh molecules coupled to a cavity for different Hamiltonians.
The spectra show the \gls{lp} and \gls{up} transitions and are calculated by Fourier transformation of the expectation value of the total dipole moment.
The different models are the molecular \gls{etc} model (pink), the molecular \gls{tc} model with only static dipole moments (blue), the molecular \gls{tc} model with only \gls{dse} contribution and the molecular \gls{tc} model (green).
The dashed line corresponds to the first vibrational resonance between the electronic states. 
All simulation were performed with cavity excitation $\omega_c = \omega_{00} = \SI{4.322}{\electronvolt}$ and coupling strength $\lambda_{c} = \SI{6.9e-3}{au}.$}
\label{fig:spec_mol_mgh}
\end{figure}
We next compare the corresponding absorption spectra for the four different molecular Hamiltonians 
of the coupled molecular-cavity system. 
In Fig.~\ref{fig:spec_mol_mgh} the spectra of the \gls{lp} and \gls{up} transitions are shown for the case of one, two and three \mgh molecules coupled to a cavity resonant with the $\ke{g,v=0}\rightarrow\ke{e,v=0}$ transition. 
The complete absorption spectra and detailed analysis of all features are provided in Section S2 of the Supporting Information. 
To obtain the spectra of the coupled molecular-cavity system, a superposition $\left(\ke{G,0}+\ke{G,1}\right)/\sqrt{2}$
was propagated for 10~ps and the expectation value of the total dipole moment was Fourier transformed.

The observed Rabi splitting of about $\approx 53$~meV is almost the same
for the four light-matter Hamiltonians, as well as for $N=1,2,3$.
Regardless of the number of molecules resonantly coupled to the cavity, the \gls{lp} and \gls{up} transitions are strongly asymmetric and redshifted with respect to $\omega_c$ and the bare molecular $\ke{g,v=0}\rightarrow\ke{e,v=0}$ transition. 
Interestingly, this asymmetry is already present when using the standard molecular \gls{tc} model Hamiltonian (Fig.~\ref{fig:spec_mol_mgh}, green). 
Including the static dipole moment leads to an increased redshift (Fig.~\ref{fig:spec_mol_mgh}, orange), 
while including only the \gls{dse} (Fig.~\ref{fig:spec_mol_mgh}, blue) leads to a decreased redshift of the \gls{lp} and \gls{up} signal. 
Consistent with the results for population dynamics, the light-matter Hamiltonian (Fig.~\ref{fig:spec_mol_mgh}, pink) is closer
to the molecular \gls{tc} model.
However, the observed differences between the different light-matter Hamiltonians are getting smaller as more molecules are included (see Figs.~\ref{fig:spec_mol_mgh}(a)-(c)).

\begin{figure}
\centering
\includegraphics[width=8.5cm]{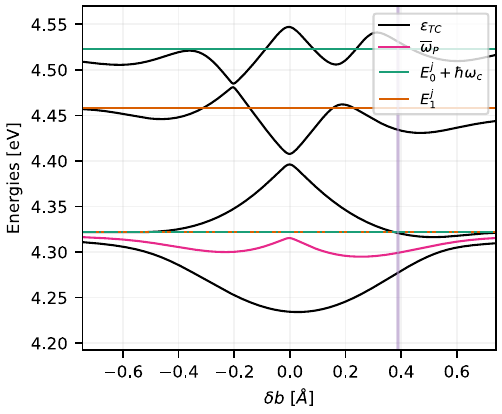}
\caption{Polaritonic eigenstates ($\epsilon_{TC}$) for the coupled system formed by the vibrational states of the single molecule electronic potentials $V_g$ and $V_e$, as a function of the relative position $\delta b$ between the two minima, for a coupling strength of $\lambda_c = \SI{6.9e-3}{au}$ and cavity frequencies of \SI{4.322}{\electronvolt}. 
The green and orange lines indicate the uncoupled vibrational eigenstates $E_i$, and the pink line indicates the average frequency $\overline{\omega}_P$ of the \gls{lp} and \gls{up} eigenstates. 
The purple vertical line indicates the natural distance between the potentials (\SI{0.388}{\AA}).}
\label{fig:eigen_pot_shift}
\end{figure}
From Fig.\ \ref{fig:spec_mol_mgh} we can conclude that the observed redshift and asymmetry in the spectrum is not caused by the \gls{dse} or the influence of the
static dipole moments, since it is already present in the molecular \gls{tc} model.
Note that similar redshifts have also been observed
in electronic structure calculations, where the effect of the electric field mode has been incorporated \cite{Schnappinger_2023}. In these cases, the redshift is a consequence of matter polarization.
In the following, we demonstrate that the observed redshift in the molecular \gls{tc} and \gls{etc} models is caused by the molecular \gls{fc} factors.
To quantify the asymmetry, we calculate the eigenvalues of the coupled system
in dependence of the relative shift in the nuclear coordinate $R$ of the \glspl{pes} for $V_g$ and $V_e$.
The relative shift between the minima of $V_g$ and $V_e$ is
defined as $\delta b$.
The resulting modified molecular Hamiltonian is coupled to the cavity mode that is resonant with the $\ke{g,v=0}\rightarrow\ke{e,v=0}$ transition to yield the molecular \gls{tc} Hamiltonian together with the transition dipole moment of \mgh.
By diagonalizing the resulting Hamiltonian, we obtain the polariton states for the shifted potential set-up as a function of $\delta b$.

The first four resulting polariton states are shown in Fig.~\ref{fig:eigen_pot_shift} (black lines).
The asymmetry of the eigenvalues around the field free transition is quantified by the average energy $\overline{\omega}_P$ of the \gls{lp} and \gls{up} states (pink lines).
The uncoupled energies are shown for reference as green and orange lines in Fig.~\ref{fig:eigen_pot_shift}.
The magnitude and asymmetry of the Rabi splitting strongly depend on the relative shift between the coupled potentials. 
Even if the potential minima are aligned ($\delta b = \SI{0}{\angstrom}$) \gls{lp} and \gls{up} are not perfectly symmetric. 
Due to the different vibrational frequencies and anharmonicity parameters of the potentials $V_g$ and $V_e$, the \gls{fc} matrix is not fully diagonal for $\delta b = \SI{0}{\angstrom}$.
For larger values of $|\delta b|$ the overlap of the wave function vanishes and effectively
decreases the transition dipole moment, resulting in a smaller Rabi splitting.
As described in Section S3 of the Support Information, this asymmetry in the Rabi splitting induced by the \gls{fc} factors
can even be observed for the case of two identical harmonic potentials. 
Higher lying vibrational states begin to mix into polariton states and thus lead to a shift in the eigenvalues.
Thus, the case of perfectly symmetric Rabi splitting in molecular \gls{esc} seems to be an exception rather than the standard case, since anharmonicity and shifted potential energy surfaces are common in molecular systems.

\section{Effective molecular Tavis-Cummings model} 

The main challenge of the molecular \gls{tc} model is the exponential
scaling of the wave function with respect to the number of molecules.
Simulating the full dynamics, including all vibrational degrees of freedom,
becomes prohibitively expensive in terms of computational effort.
Simplifying the description of the matter and replacing the molecules with effective \glspl{tls} can greatly reduce the computational cost.

In the following, we derive an effective model based on an ensemble of \glspl{tls} coupled to a single cavity mode, starting from the molecular \gls{etc} Hamiltonian after the \gls{cs} transformation. The effect of the static dipole moments and the
\gls{dse} is thus preserved.
Each molecule in the ensemble is replaced by a two-level emitter defined by the two electronic states. 
The matter Hamiltonian shown in Eq.~\eqref{eq:Hm_1mol} is simplified to $H_M = \omega_{eg}\hat\sigma^\dagger\hat\sigma$, where $\omega_{eg}$ is the energy difference between the $\ke{g,v=0}\rightarrow\ke{e,v=0}$ transition. 
The nuclear position-dependent dipole moment and dipole moment squared operators $\tilde{\mu}$ and $\tilde{\mu}^2$ of each molecule are replaced by the corresponding expectation values $\ev{\tilde\mu}$ and $\ev{\tilde\mu^2}$ at the \gls{fc} point for each of the electronic states. 
A detailed derivation of the general \gls{tls} model Hamiltonian and all of its coupling terms can be found in Section S4 of the Supporting Information.

\begin{figure}
    \centering
    \includegraphics[width=8.5cm]{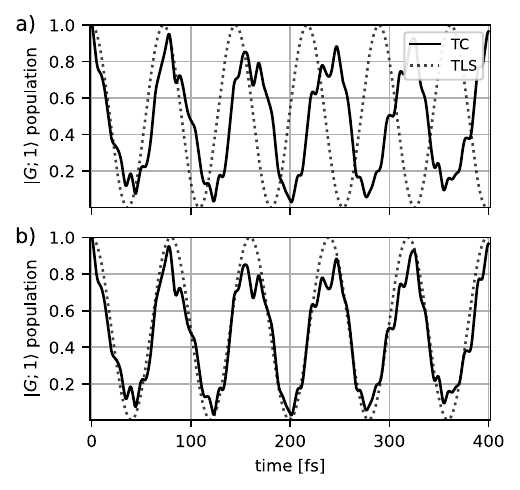}
    \caption{Comparison of propagation of a single molecule system to a single \gls{tls} using the \gls{tc} model a) with the same cavity parameters: coupling strength of $\lambda_c=\SI{6.9e-3}{au}$, cavity resonant to the \gls{tls} frequency excitation $\omega_{eg}^{TLS} = 4.322$~eV, b) and with optimized \gls{tls} parameters: coupling strength of $\lambda_c=\SI{6.25e-3}{au}$,  levels resonance $\omega_{eg}^{TLS} = 4.297$~eV, and cavity detuning of \SI{-3.59}{\meV}}
    \label{fig:Mol_TLS_sameparameters}
\end{figure}
To evaluate the validity of the \gls{tls} approximation, we compare the dynamics of the $\ke{G;1}$ population in the case of one \mgh molecule coupled to a cavity with the results obtained for a \gls{tls} using the \gls{tc} model for both. 
The results
are shown in Fig.~\ref{fig:Mol_TLS_sameparameters}. 
If the resonant frequency of the cavity and the energy difference of the levels are not modified ($\omega_c = \omega_{eg}^{TLS} = \SI{4.322}{\electronvolt}$) the temporal evolution of the $\ke{G;1}$ state population is different (see Fig.~\ref{fig:Mol_TLS_sameparameters}~a)). 
The two notable differences are an increase in Rabi frequency for the \gls{tls} approximation and the absence of fine structure in the oscillations caused by
the vibrational motion.
The underlying reason for this discrepancy can be found in the construction of the \gls{tls} model: the absence of vibrational degrees of freedom and a non-diagonal \gls{fc} matrix results in a symmetric Rabi splitting (see previous section).
Thus, the energetics as well as the population dynamics are different in the \gls{tls} model and in the molecular system when the same cavity parameters are used. 
To improve the \gls{tls} model, we optimize its parameters to mimic the energetics of the molecular polaritonic states. 
These optimized parameters can be obtained by fitting the \gls{tls} polariton energies to the molecular absorption spectra. A detailed explanation of this optimization process can be found in Section S4 of the Supporting Information. 
The population dynamics of the $\ke{G;1}$ state using the optimized \gls{tls} parameter ($\omega_{eg}^{TLS} = \SI{4.297}{\electronvolt}$ and cavity detuning of \SI{-3.59}{\meV}) compared to the molecular simulation is shown in Fig.~\ref{fig:Mol_TLS_sameparameters}~b). 
With the optimized parameters, the \gls{tls} model qualitatively reproduces the population dynamics of the molecular system and exhibits an identical Rabi frequency.

\begin{figure}
\centering
\includegraphics[width=8.5cm]{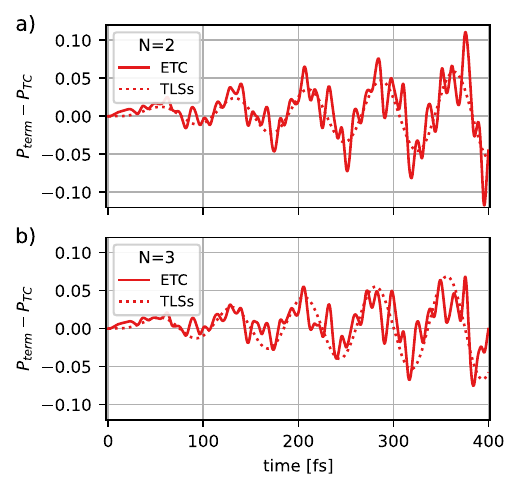}  
\caption{State $\ke{G;1}$ population dynamics of the optimized \glspl{tls} model and the molecular simulations for a) two and b) three \mgh molecules coupled to the cavity mode, comparing the full-\glspl{tls} and \gls{etc} model Hamiltonian population differences to the corresponding \glspl{tls} or molecular \gls{tc} model.}
\label{fig:tls_opt_comparison}
\end{figure}
To extend the validity of the \gls{tls} approximation, we compare the population difference between the \gls{tc} and the extended models in the molecular
and the optimized \glspl{tls} case. 
The results for two and three \mgh molecules/emitters are shown in Fig.~\ref{fig:tls_opt_comparison}.
The qualitative agreement of the \gls{tls} improves as the number of molecules increases. 
Similarly to section~\ref{sec:pop_mol}, the influence of nuclear wave packet dynamics becomes smaller with increasing number of molecules, since the photonic excitation is evenly distributed over more molecules, resulting in less vibrational excitation per molecule. 
It should be noted that the optimized parameters of the \gls{tls} models are slightly different for one, two, and three \mgh molecules (see Table S1 in the Supporting Information). However, we observe a tendency to values close to the original molecular parameters as the number of molecules increases.

\begin{figure}
    \centering
    \includegraphics[width=8.5cm]{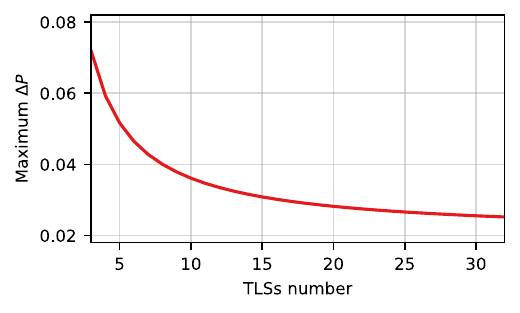}
    \caption{Maximum $\ke{G;1}$ population difference between the \gls{tls}-\gls{etc} model Hamiltonian and the \gls{tc} model Hamiltonian after \SI{400}{\femto\second} as function of the number of \glspl{tls} in the ensemble.}
    \label{fig:popdiff_largeNTLS}
\end{figure}
Based on these results, we assume that our \gls{tls} model with the optimized parameters for 3 \mgh molecules is capable of capturing the essential features of the coupled system dynamics for the general $N$ molecules situation. 
The \gls{tls} model allows us to simulate larger systems and we choose a maximum of 33 emitters to demonstrate convergence. 
In Fig.~\ref{fig:popdiff_largeNTLS} the maximum deviation of the population difference between the \gls{tls}-\gls{etc} and \gls{tc} model Hamiltonians
within \SI{400}{\femto\second} is plotted as a function of the \gls{tls} number $N$. This maximum deviation is given by
\begin{equation}
    \max\Delta P (t, N) = \max\left(|P_\text{ETC} - P_\text{TC}|\right),
\end{equation}
where $P_\text{ETC}$ and $P_\text{TC}$ are the state $\ke{G;1}$ populations for the different Hamiltonian propagations. Their maximum difference is calculated by fitting the local maxima to a linear regression and deriving its value at time $t$. The maximum deviation in the population decreases with increasing number of \glspl{tls} coupled to the cavity mode, as shown in Fig.~\ref{fig:popdiff_largeNTLS}. 
An increase in $N$ leads to an increase in the energy shifts of the states and couplings between excited states in Eq.~\eqref{H_Nmol_matrix}. 
However, due to rescaling of the coupling strength, this increase is fully compensated and, as a consequence, the difference between the \gls{tls}-\gls{etc} and \gls{tc} model Hamiltonians is inversely proportional to $N$.
To confirm this behavior, a function $\max \Delta P = 0.020 + 0.151~N^{-0.970}$
has been fitted to the calculated curve in Fig.~\ref{fig:popdiff_largeNTLS}.
This fit confirms that the maximum difference decays with an exponent of -0.970
and converges to a finite difference of 0.020 for large $N$.

\begin{figure}
\centering
 \includegraphics[width=8.5cm]{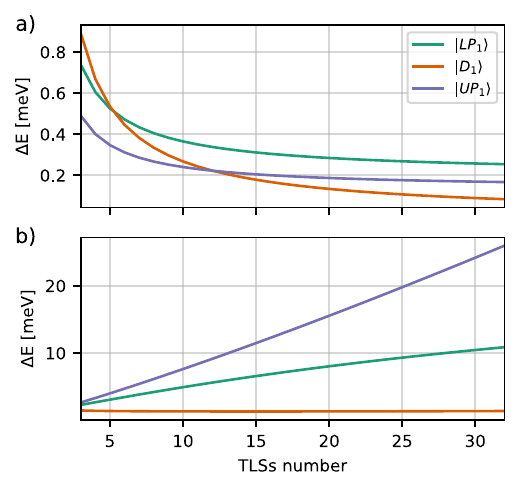} 
\caption{\gls{tls} eigenvalues differences between diagonalized \gls{tls}-\gls{etc} and \gls{tc} models as functions of the system size for a) the \gls{cs} transformed and b) non \gls{cs} transformed systems. Scaled coupling strength of $\lambda_c = \SI{6.52e-3}{au}$, level frequency difference of $\omega_{eg}=\SI{4.312}{eV}$, and cavity detuning by \SI{11.6}{meV}. \textit{UP}$_1$ stands for the upper polariton of the first excitation manifold, \textit{ LP}$_1$ the lower polariton, and \textit{ D}$_1$ the dark states. }
\label{fig:energyshift_N}
\end{figure}
In addition to studying the population dynamics, we examine the eigenenergies of the \gls{tls}-\gls{etc} model Hamiltonian to determine the influence of the \gls{cs} transformation on the polaritonic states, as well as the effect of including static dipole moments and \gls{dse} terms. 
By diagonalizing the $N$-\glspl{tls} Hamiltonian, we obtain two polariton states $\ke{LP}$, $\ke{UP}$, and $N-1$ dark states summarized as $\ke{D_1}$.
The variation in the eigenergies of the \gls{tls}-\gls{etc} and \gls{tc} model Hamiltonians are given by
\begin{equation}
\Delta E_S = \ev{\hat{H}^\text{TLS}_\text{ETC}}_S - \ev{\hat{H}^\text{TLS}_\text{TC}}_S,
\end{equation} 
where $S$ is a particular polariton eigenstate of the respective Hamiltonian.
In Fig.~\ref{fig:energyshift_N}(a) these energy differences between the eigenergies of the \gls{tls}-\gls{etc} and \gls{tc} model Hamiltonians are shown as a function of the \gls{tls} number.
All three curves, upper and lower polarition state, as well as the dark states
have also been fitted to the following functions:
\begin{eqnarray*}
\Delta E_\text{LP} = 0.203\,\text{meV} + 1.595\,\text{meV}~N^{-0.995} \\
\Delta E_\text{D~} = 0.000\,\text{meV} + 2.672\,\text{meV}~N^{-1.000} \\
\Delta E_\text{UP} = 0.134\,\text{meV} + 1.077\,\text{meV}~N^{-1.008}.
\end{eqnarray*}
The energy differences of all states are approximately inversely proportional to the number of \glspl{tls}, similar to the trend observed in Fig.~\ref{fig:popdiff_largeNTLS}. This is consistent with a $\lambda^2$ square scaling as it appears in the \gls{dse} term.
The difference between the \gls{tls}-\gls{etc} and \gls{tc} model Hamiltonians converges to zero for the dark states.
However, for the \gls{lp} state and the \gls{up} state it converges for large $N$ to finite values of 0.203\,meV and 0.134\,meV respectively. 
This trend is consistent with the redshift and asymmetry observed in the spectra in Fig.~\ref{fig:spec_mol_mgh}. 
The covariance matrices for the fitted curves of Maximum $\Delta P$ and $\Delta E_S$ are shown in Section S4 of the Supporting Information.

Figure \ref{fig:energyshift_N}~b) shows the energy difference for \gls{tls} without the \gls{cs} transformation. The energies of the two bright polaritonic states diverge for an increase in $N$.
Note that similar to Fig.~\ref{fig:pop_dyn_multi_mgh}(b) and (d) the \gls{tls} represent molecules that are aligned in parallel. This results in an increasing total dipole moment with increasing $N$ and causes divergent behavior.
The dark state energies are not affected by the \gls{cs} transformation, as they are decoupled from the cavity mode and have no photonic contribution.

\section{Conclusion}
We have extended the molecular \gls{tc} model
to include both state-specific static dipole moments and the \gls{dse} contribution in the light-matter Hamiltonian. 
Starting from the non-relativistic Pauli-Fierz Hamiltonian, we derived a light-matter Hamiltonian in the \gls{cs} basis and used the \gls{rwa} to describe molecules under \gls{esc}. 
The studied molecular system consists of a varying number of \mgh molecules coupled to a single-photon mode of an optical cavity resonant with the first electronic transition in \mgh. 
By analyzing the difference in the population dynamics obtained with the molecular \gls{tc} model and the generalized molecular \gls{etc}, we could identify changes independent of the number of molecule.
The deviations caused by either the static dipole moments or the \gls{dse} contributions are significantly different and give rise to the overall difference in the molecular \gls{etc} model.
Thus, both components are essential to describe molecules coupled to a cavity. 
In line with the literature~\cite{Schafer2020-cb,Sidler2022,Sidler2024,Schnappinger2023-hh,Ruggenthaler2023,Schnappinger_2023,Horak2024-cd}, we can therefore emphasize that the \gls{dse} should not be neglected, even in the \gls{esc} regime. Although the discrepancy between the molecular \gls{tc} and \gls{etc}
decreases for larger $N$, we could show that it convergences to finite values.
Another aspect that has been discussed in the literature, mainly for ab initio methods~\cite{Haugland2020-xh,Foley-2023,Castagnola2023-tb} is the use of the \gls{cs} transformation for systems with a static dipole moment. 
We could show that the \gls{cs} transformation becomes relevant when the total dipole moment of the molecular ensemble is nonzero. 

The \gls{dse} contribution itself does not depend on the number of photons, and therefore should be affected only indirectly by the photon loss of the cavity. However, photon decay and dissipation can affect the dynamics of molecular systems under \gls{esc} quite drastically~\cite{Davidsson2020-qb,Davidsson2023-xa}. Thus it would be an interesting next step to study the interplay of \gls{dse} and photon loss.

Analysis of polaritonic absorption spectra of coupled \mgh-cavity systems
revealed that \gls{lp} and \gls{up} are strongly redshifted and are asymmetric
in intensity. This phenomenon is independent of the exact model used and could even be observed for harmonic potentials.
We could identify this shift as a result of the molecular \gls{fc} factors, which
depended on the relative change in the equilibrium bond length between the ground and excited state.
In the \gls{vsc} regime, such a redshift is associated with a self-consistent treatment of the electronic structure problem~\cite{Sidler2024,Schnappinger_2023,Horak2024-cd}, which is not the case in our \gls{esc} simulation. 

Furthermore, we investigated the possibility of representing the molecules
as \glspl{tls} with only two electronic states each and without nuclear degrees of
freedom. We constructed an ensemble of \glspl{tls} to estimate the population dynamics of larger ensembles of up to 33 \mgh molecules based on the presented molecular \gls{etc} model. Such a model could be used to describe larger ensembles where the nuclear motion is not important and does not lead to reactions or nuclear rearrangements.

\begin{acknowledgments}
This project has received funding from the European Research Council (ERC) under the European Union’s Horizon 2020 research and innovation program (grant agreement no. 852286).
\end{acknowledgments}

\section*{Supplementary Material}
See the supplementary material for the full derivation of the extended molecular Tavis-Cummings (ETC) ansatz as well as the effective Tavis-Cummings model. In addition, a detailed analysis of the polaritonic absorption spectra is given.   

\section*{Author Declaration Section}
\subsection*{Conflict of Interest Statement }
The authors have no conflicts to disclose

\subsection*{Author Contributions}
\textbf{Lucas Borges}: Data curation (lead); Formal analysis (equal); Methodology (equal); Visualization (lead); Writing – original draft (equal). \textbf{Thomas Schnappinger}: Conceptualization (equal); Formal analysis (equal); Methodology (equal); Investigation (equal); Writing – original draft (equal), Writing – review \& editing (equal). \textbf{Markus Kowalewski}: Conceptualization (equal); Formal analysis (equal); Funding
acquisition (lead); Methodology (equal);
Project administration (lead); Supervision (lead); Writing – original draft (equal); Writing – review \& editing (equal).

\section*{Data Availability}
The data that support the findings of this study are available from the corresponding author upon reasonable request.

\bibliography{lit.bib}

\end{document}


\title{Supporting Information: \\Extending the Tavis-Cummings model for molecular ensembles -- Exploring the effects of dipole self energies \\and static dipole moments}
\author{Lucas Borges}
\affiliation{Department of Physics, Stockholm University, AlbaNova University Center, SE-106 91 Stockholm, Sweden}

\author{Thomas Schnappinger}
\email{thomas.schnappinger@fysik.su.se}
\affiliation{Department of Physics, Stockholm University, AlbaNova University Center, SE-106 91 Stockholm, Sweden}

\author{Markus Kowalewski}
\email{markus.kowalewski@fysik.su.se}
\affiliation{Department of Physics, Stockholm University, AlbaNova University Center, SE-106 91 Stockholm, Sweden}

\maketitle

\clearpage
\tableofcontents
\clearpage

\section{The generalized molecular Tavis-Cummings model}

After the length gauge and the \gls{cs} transformation, the interaction part of the Pauli-Fierz Hamiltonian is given by:
\begin{equation}
\hat{H}_{int} = -\sqrt{\frac{\omega_c}{2}}(\hat{a}^\dagger + \hat{a})\left(\lambda \Tilde{\mu}\right) + \frac{1}{2} \left(\lambda\Tilde{\mu}\right)^2,
\label{SI:int}
\end{equation}
The total dipole moment operator $\tilde{\mu}$ of a ensemble of $N$ molecules after the \gls{cs} transformation reads:
\begin{equation}\label{SI:mu}
\tilde{\mu} = \sum_{i=1}^{N} \hat{\mu}^{(i)} - \expval{\hat{\mu}}_0,
\end{equation}
where $\expval{\hat{\mu}}_0$ is the ground state permanent dipole moment of the whole ensemble and $\hat{\mu}^{(i)}$ is the dipole moment operator of the individual molecule in the nuclear subspace:
\begin{equation}
    \hat{\mu}^{(i)} = \mu_{gg}\hat{\sigma}^{(i)}\hat{\sigma}^{(i)\dagger} + 
\mu_{ee}\hat{\sigma}^{(i)\dagger}\hat{\sigma}^{(i)} + \mu_{eg}\left( \hat{\sigma}^{(i)} + \hat{\sigma}^{(i)\dagger} \right)\,,    
\end{equation}
where $\mu_{mn}\equiv\ev{\mu}_{mn}(\bm{R_i})$ are the $\bm{R_i}$ dependent dipole matrix elements between electronic states $m$ and $n$ respectively. The corresponding total squared dipole operator is given by
\begin{equation}
\label{eq:mu2_2m}
   \tilde{\mu}^2 = \sum_{i=1}^{N} \left(\hat{\mu}^{(i)}\right)^2 +
\hat{\mu}^{(i)}\left( \sum_{j \neq i}^{N} \hat{\mu}^{(j)} - 2\expval{\hat{\mu}}_0\right)
 + \expval{\hat{\mu}}_0^2.
\end{equation}
Here $\left(\hat{\mu}^{(i)}\right)^2$ is the squared dipole operator of the individual molecule in the nuclear subspace:
\begin{equation}\label{SI:mu2}
\left(\hat{\mu}^{(i)}\right)^2 =  
\mu^2_{gg}\hat{\sigma}^{(i)}\hat{\sigma}^{(i)\dagger}  + \mu^2_{ee}\hat{\sigma}^{(i)\dagger}\hat{\sigma}^{(i)} +\mu^2_{eg}\left( \hat{\sigma}^{(i)} + \hat{\sigma}^{(i)\dagger} \right)\,,    \end{equation}
where $\mu^2_{mn}\equiv\ev{\mu^2}_{mn}(\bm{R_i})$ are the squared dipole moments dependent on $\bm{R_i}$ between electronic states $m$ and $n$ respectively.

In the following derivation of the coupling terms in the generalized molecular Tavis-Cummings model we take advantage of the fact that the individual molecular wave functions $\ket{g^{(i)}}$ and $\ket{e^{(i)}}$ as well as the bare Fock states are orthonormal and the individual molecules are non-interacting. For brevity, prefactors are omitted, and the same color schema as in the manuscript is used to highlight relevant contributions. 
In the bare-state basis truncated to a maximum of two excitations, the first term in Eq.~\eqref{SI:int} leads to linear dipole coupling terms between states in which the photonic excitation, i.e. the Fock state, changes. The first type of linear coupling term connects different molecular ensemble states:
\begin{align}
\bra{G;0}  (\hat{a}^\dagger + \hat{a})\tilde\mu \ket{E^{(a)};1} & =
\bra{g^{(a)}} \hat{\mu}^{(a)} \ket{e^{(a)}} + \braket{g^{(a)}}{e^{(a)}} \left(\dots\right) = \mu_{eg}^{(a)}
\label{eq:lin1}\\
\boxthistle{$\bra{G;1} (\hat{a}^\dagger + \hat{a})\tilde\mu \ket{E^{(a)};0}$} & =
\bra{g^{(a)}} \hat{\mu}^{(a)} \ket{e^{(a)}} + \braket{g^{(a)}}{e^{(a)}} \left(\dots\right) = \mu_{eg}^{(a)} \label{eq:lin2}\\
\boxthistle{$\bra{G;2} (\hat{a}^\dagger + \hat{a}) \tilde\mu \ket{E^{(a)};1}$} & =
\bra{g^{(a)}} \hat{\mu}^{(a)} \ket{e^{(a)}} + \braket{g^{(a)}}{e^{(a)}} \left(\dots\right) = \mu_{eg}^{(a)} \label{eq:lin3}\\
\boxthistle{$\bra{E^{(a)};1} (\hat{a}^\dagger + \hat{a})\tilde\mu \ket{\mathcal{E}^{(a,b)};0}$} & =
\bra{g^{(b)}} \hat{\mu}^{(b)} \ket{e^{(b)}} + \braket{g^{(b)}}{e^{(b)}} \left(\dots\right) = \mu_{eg}^{(b)} \label{eq:lin4}
\end{align}
The second type of liner dipole interactions couple different vibrational states within the same electronic state:
\begin{equation}
    \boxygreen{$\bra{E^{(a)};1} (\hat{a}^\dagger + \hat{a})\tilde\mu \ket{E^{(a)};0}$}  = \mu_{ee}^{(a)}+ \sum_{b \neq a}^{N} \mu_{gg}^{(b)}-\expval{\hat{\mu}}_0\label{eq:lin5}
\end{equation}
This type of coupling term is zero for all states formed by the ensemble ground state due to the \gls{cs} transformation. The coupling terms shown in Eqs.~\ref{eq:lin1} and~\ref{eq:lin5} are not part of the standard \gls{tc} Hamiltonian. 

The second term in Eq.~\eqref{SI:int} gives rise to \gls{dse} terms between ensemble states where the photonic excitation is not changing. These terms can be divided into three groups. The first group of terms provides state-specific energy shifts for the three types of molecular ensemble states:
\begin{align}
\bra{G;n} \tilde\mu^2 \ket{G;n}  = &\sum_{a=1}^{N} \left(\mu_{gg}^2\right)^{(a)} + \mu_{gg}^{(a)}\left( \sum_{b \neq a}^{N} \mu_{gg}^{(b)} - 2\expval{\hat{\mu}}_0\right) + \expval{\hat{\mu}}_0^2 \label{eq:dse1_dia}\\
\bra{E^{(a)};n} \tilde\mu^2 \ket{E^{(a)};n}  = & \left(\mu_{ee}^2\right)^{(a)} +  \mu_{ee}^{(a)} \left( \sum_{b=1}^{N-1} \mu_{gg}^{(b)} - 2\expval{\hat{\mu}}_0\right) + \expval{\hat{\mu}}_0^2 \nonumber \\
 & + \sum_{b=1}^{N-1} \left(\mu_{gg}^2\right)^{(b)} + \mu_{gg}^{(b)}\left( \sum_{c \neq b}^{N-1} \mu_{gg}^{(c)} - 2\expval{\hat{\mu}}_0\right) \label{eq:dse2_dia}\\
 \bra{\mathcal{E}^{(a,b)};n} \tilde\mu^2 \ket{\mathcal{E}^{(a,b)};n}  = & \left(\mu_{ee}^2\right)^{(a)} +  \left(\mu_{ee}^2\right)^{(b)} 
 + 2 \mu_{ee}^{(a)} \mu_{ee}^{(b)} + \expval{\hat{\mu}}_0^2  \nonumber \\
 & + \left( \mu_{ee}^{(a)} +  \mu_{ee}^{(b)}  \right) \left( \sum_{c=1}^{N-2} \mu_{gg}^{(c)} - 2\expval{\hat{\mu}}_0\right)   \nonumber \\
  & + \sum_{c=1}^{N-2} \left(\mu_{gg}^2\right)^{(c)} + \mu_{gg}^{(c)}\left( \sum_{d \neq c}^{N-2} \mu_{gg}^{(d)} - 2\expval{\hat{\mu}}_0\right) \label{eq:dse3_dia}
\end{align}
The second group of \gls{dse} contributions couples state which share the same type of  molecular ensemble states.
\begin{align}
\boxyorange{$ \bra{E^{(a)};n} \tilde\mu^2 \ket{E^{(b)};n}$}  = & \mu_{eg}^{(a)}\mu_{eg}^{(b)} \label{eq:dse1_od_s}\\
\boxyorange{$\bra{\mathcal{E}^{(a,b)};n} \tilde\mu^2 \ket{\mathcal{E}^{(b,c)};n}$}  = & \mu_{eg}^{(a)}\mu_{eg}^{(c)}\label{eq:dse2_od_s}\\
\bra{\mathcal{E}^{(a,b)};n} \tilde\mu^2 \ket{\mathcal{E}^{(c,d)};n}  = & 0 \label{eq:dse3_od_s}
\end{align}
The last group of \gls{dse} coupling connects states which share different types of molecular ensemble states.
\begin{align}
\bra{G;n} \tilde\mu^2 \ket{E^{(a)};n}  = & \left(\mu_{eg}^2\right)^{(a)} +  \mu_{eg}^{(a)} \left( \sum_{b=1}^{N-1} \mu_{gg}^{(b)} - 2\expval{\hat{\mu}}_0\right) \label{eq:dse1_od_d}\\ 
\bra{G;n} \tilde\mu^2 \ket{\mathcal{E}^{(a,b)};n}  =  & \mu_{eg}^{(a)}\mu_{eg}^{(b)} \label{eq:dse2_od_d}\\ 
\bra{E^{(a)};n} \tilde\mu^2 \ket{\mathcal{E}^{(a,b)};n}  =  & \left(\mu_{eg}^2\right)^{(b)} + \mu_{ee}^{(a)} \mu_{eg}^{(b)} + \mu_{eg}^{(b)} \left( \sum_{c=1}^{N-2} \mu_{gg}^{(c)} - 2\expval{\hat{\mu}}_0\right)
\label{eq:dse3_od_d}\\ 
\bra{E^{(a)};n} \tilde\mu^2 \ket{\mathcal{E}^{(b,c)};n}  =  & 0 \label{eq:dse4_od_d}
\end{align}

To reduce the complexity of the interaction Hamiltonian in the extended molecular Tavis-Cummings model, we apply the commonly used \gls{rwa}~\cite{QOiPS_20001}. This approximation affects all coupling terms whose contributions to the system dynamics are negligible. In the near-resonance regime, the liner dipole coupling terms of Eq.~\eqref{eq:lin1} oscillate at twice the field frequency. Thus, the contribution of the coupling to the dynamics of the system is negligible. This argument does not hold for the coupling between states of the same matter excitation of Eq.~\eqref{eq:lin5} arising from the permanent dipole moment component, since its time evolution in the interaction picture is driven by the optical field frequency. The squared transition dipole moment couplings of Eqs.~\eqref{eq:dse1_od_d},~\eqref{eq:dse2_od_d} and~\eqref{eq:dse3_od_d} have a negligible influence on the system dynamics due to the large energy difference between the coupled states together with the squared coupling parameter $\lambda^2$.

To verify the \gls{rwa} we determine the influence of the neglected terms on the population dynamics for a single \mgh molecule coupled to a cavity. The \gls{etc} Hamiltonian of single molecule coupled to single cavity  mode in the \gls{rwa} reads:
\begin{align}    
\hat{H} = \hat{H}_M &+ \omega_c \left(\hat{a}^{\dagger} \hat{a} + \frac{1}{2} \right) -\sqrt{\frac{\omega_c}{2}}\lambda \mu_{eg} \left(\hat{a}\hat{\sigma}^\dagger  + \hat{a}^\dagger\hat{\sigma}\right) -\sqrt{\frac{\omega_c}{2}}\lambda\left(\hat{a} + \hat{a}^\dagger\right) \left(\mu_{ee}- \expval{\hat{\mu}}_0\right) \hat{\sigma}^\dagger \hat{\sigma}\nonumber\\ & +  \frac{1}{2}
\lambda^2 \left(\mu_{gg}^2- \expval{\hat{\mu}}_0^2\right)\hat{\sigma}\hat{\sigma}^\dagger +
\frac{1}{2}
\lambda^2\left(\mu_{ee}^2- 2\mu_{ee}\expval{\hat{\mu}}_0 + \expval{\hat{\mu}}_0^2\right)\hat{\sigma}^\dagger\hat{\sigma}. 
\end{align}\label{eq:1molHamilt}

\begin{figure}[htb]
\centering
\includegraphics[width=.95\textwidth]{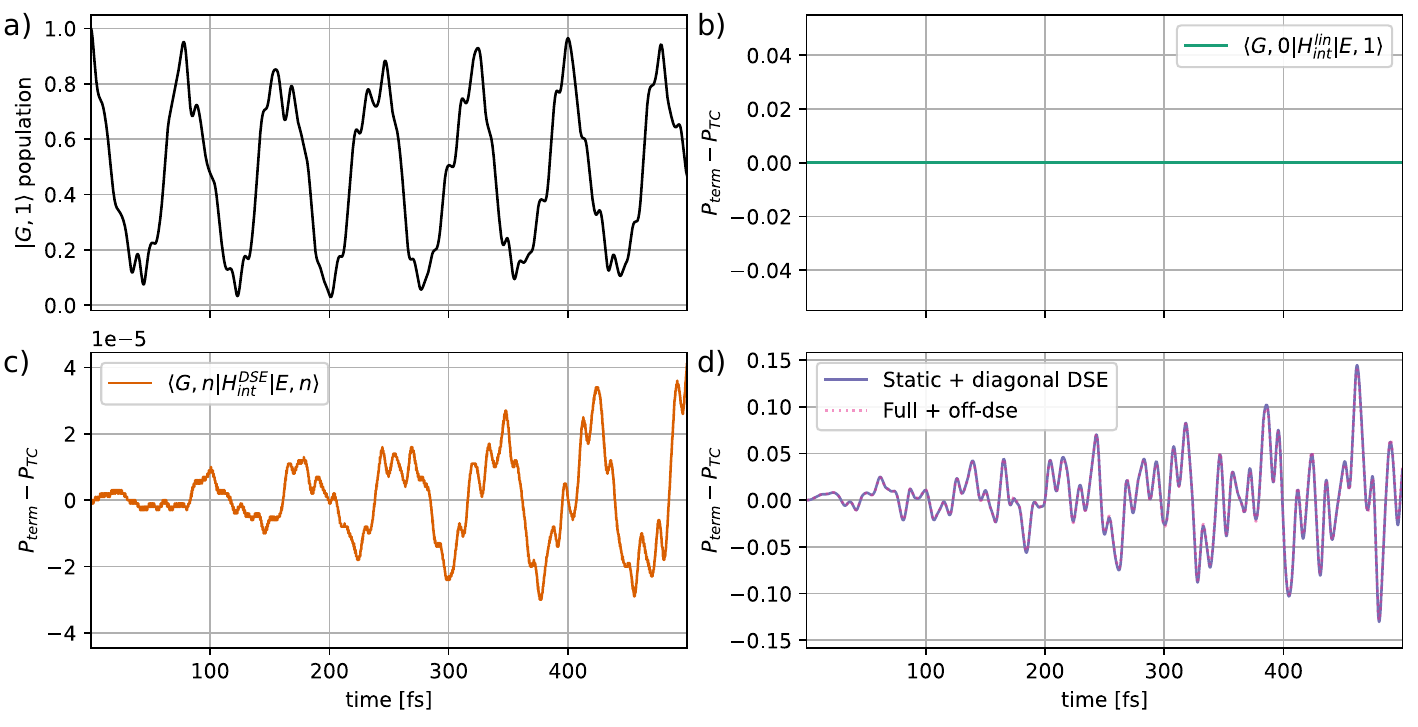}
\caption{a) Population dynamics of state $\ket{g,1}$ for a single \mgh molecule coupled to a cavity obtained using the standard molecular \gls{jc} Hamiltonian. b) Difference in the population dynamics with respect to the \gls{jc} results if only the counter-rotation coupling Eq.~\eqref{eq:lin1} is included. c) Population difference if only the \gls{dse} coupling (see Eq.~\eqref{eq:dse1_od_d}) is included. d) Population differences of the \gls{etc} dynamics with the \gls{rwa} (purple) and without the \gls{rwa} (pink). The cavity frequency  is \SI{4.322}{\electronvolt} and a coupling strength of \SI{6.9e-3}{au} is used.}
\label{fig:benchmark}
\end{figure}

The population dynamics of the $\ket{g,1}$ state obtained using the standard molecular \gls{jc} model, i.e. including only the transition dipole moment coupling, see Eq.~\eqref{eq:lin2}, is plotted in Fig.~\ref{fig:benchmark}~a) and used as a reference in the following. If only the additional linear dipole coupling term is included, see Eq.~\eqref{eq:lin1}, the difference in population dynamics with respect to the molecular \gls{jc} result is practically zero for the coupling strength used, see Fig.~\ref{fig:benchmark}~b). In the case of a single molecule, only the \gls{dse} term shown in Eq.~\eqref{eq:dse1_od_d} exists and is therefore affected by the \gls{rwa}. Its influence on the $\ket{g,1}$ dynamics with respect to the molecular \gls{jc} population is very small, as shown in Fig.~\ref{fig:benchmark}~c). To investigate the influence of the \gls{rwa} on the \gls{etc} model, we compare the population difference with respect to the molecular \gls{jc} result obtained with the \gls{rwa} (purple curve in Fig.~\ref{fig:benchmark}~d)) and without the \gls{rwa} (pink curve in Fig.~\ref{fig:benchmark}~d)). The small observed differences indicate that the \gls{rwa} can be safely applied and that the contributions of Eqs.~\eqref{eq:lin1},~\eqref{eq:dse1_od_d},~\eqref{eq:dse2_od_d} and~\eqref{eq:dse3_od_d} can be neglected.

\begin{figure}[htb]
    \centering
    \includegraphics[width=.95\textwidth]{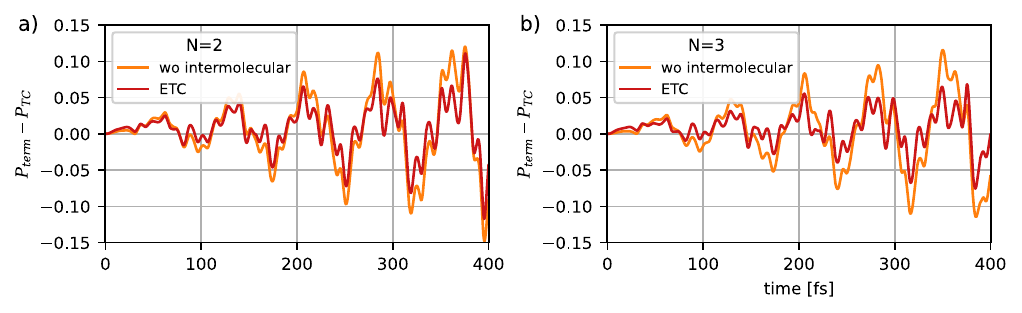}
    \caption{Difference in the population dynamics of state $\ket{G;1}$ with respect to the molecular \gls{tc} model results for a) two and b) three molecules, considering the \gls{etc} model (red curve) and without the intermolecular dipole-dipole coupling (orange curves).}
    \label{fig:wo_intermolecular}
\end{figure}

To understand the effects of the intermolecular dipole-dipole term arising from the \gls{dse} contribution of Eq.~\eqref{eq:dse1_od_s}, we propagated the Hamiltonians without its contribution. The differences in the propagation with respect to the molecular \gls{tc} is shown in Fig.~\ref{fig:wo_intermolecular}. The presence of this coupling in the \gls{etc} model decreases the population difference in the case of $N>1$ molecules.

\begin{figure}[htb]
    \centering
    \includegraphics[width=8.5cm]{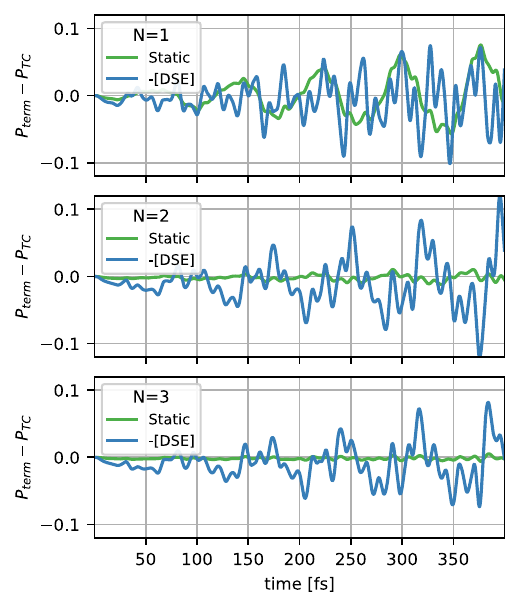}
    \caption{Difference in the population dynamics of state $\ket{G;1}$ with respect to the molecular \gls{tc} model results for one, two and three molecules, including the static dipole moment and \gls{dse} coupling terms separately in the molecular \gls{tc} model Hamiltonian. }
    \label{fig:popdiff}
\end{figure}

In Fig.~\ref{fig:popdiff} we plot the variation in the population of state $\ket{G,1}$ for the cases where the molecular \gls{tc} model is extended with only the static dipole moment coupling or the \gls{dse} coupling terms. 
We note that with the increase of molecules in our model the population difference of the static dipole moment contribution is reduced, while the \gls{dse} becomes more dominant in the complete \gls{etc} model. 

\clearpage
\section{\label{sec:spectra}Absorption Spectra}

\begin{figure}[htb]
\centering
\includegraphics[width=\textwidth]{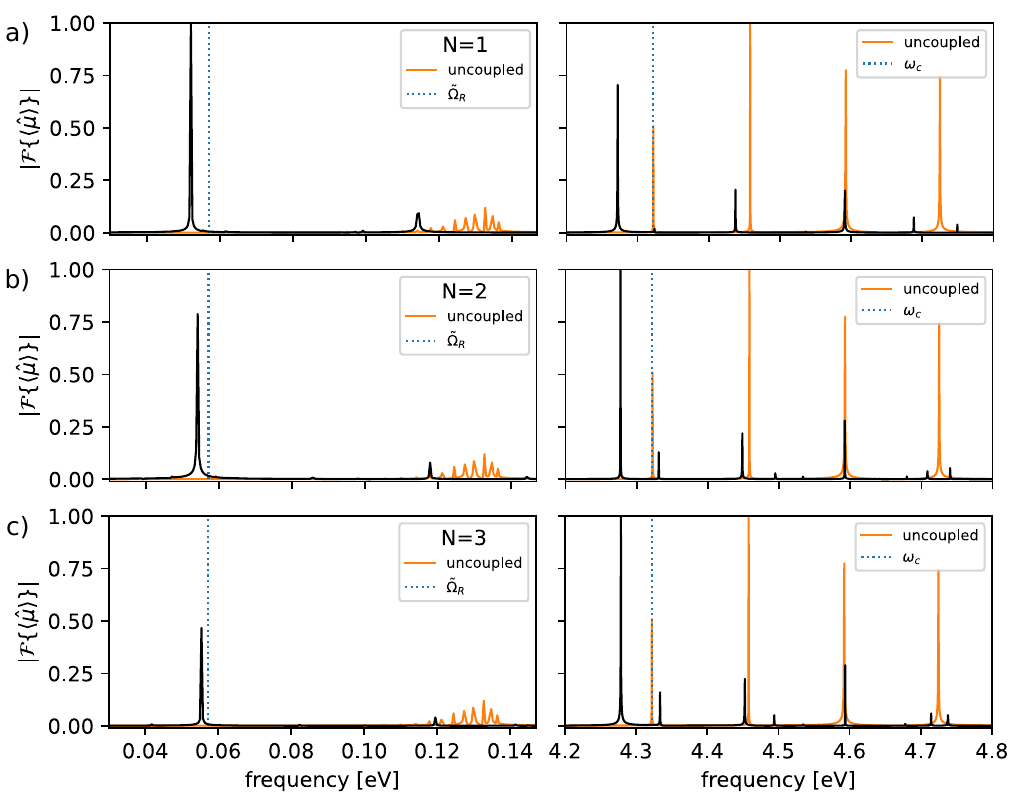}
\caption{Polaritonic absorption spectra for a) one, b) two and c) three molecule considering the \gls{etc} model with coupling strength $\sqrt{N}\lambda_c = \SI{6.9e-3}{au}$ and cavity excitation frequency $\omega_c=4.322$~eV (black curve). The orange curve corresponds to the uncoupled case. The left column shows the IR part of the spectra, the right column shows the relevant part of the UV spectra. The blue dotted lines in the left column correspond to the Rabi frequency of the equivalent \gls{tls} model $\tilde\Omega_R$, and in the right column, they are indicating the cavity frequency $\omega_c$.}
\label{fig:1mol_abspectra}
\end{figure}

The absorption spectra for the coupled molecular-cavity system were calculated by Fourier transform of the time-dependent expectation value of the total dipole moment. The resulting absorption spectra contain both vibrational and electronic transitions in the coupled molecule-cavity systems. In Fig.~\ref{fig:1mol_abspectra} these spectral regions are shown for one, two, and three \mgh molecules coupled to a cavity (black) and without cavity coupling (orange). The transition between the \gls{lp} state and the \gls{up} state is the most prominent feature in the IR spectra, left column of Fig.~\ref{fig:1mol_abspectra}. The observed Rabi frequency $\Omega_R$ of $52.1\,meV$, for the single molecule case, is slightly redshifted compared to the approximate value of $\sqrt{2\omega_c}\lambda\mu_{eg} = 57\,meV$. The UV spectra (right column of Fig.~\ref{fig:1mol_abspectra}) show the formation of the \gls{lp} and \gls{up} states, which are strongly red-shifted with respect to the cavity frequency $\omega_c$. Within the vibrational progression, peaks are shifted, and additional splittings due to the light-matter interaction are visible, e.g. around 4.7\,eV.

\clearpage
\section{\label{sec:PFC}Polaritonic Franck-Condon Factors} 

\begin{figure}[htb]
\centering 
\includegraphics[width=\textwidth]{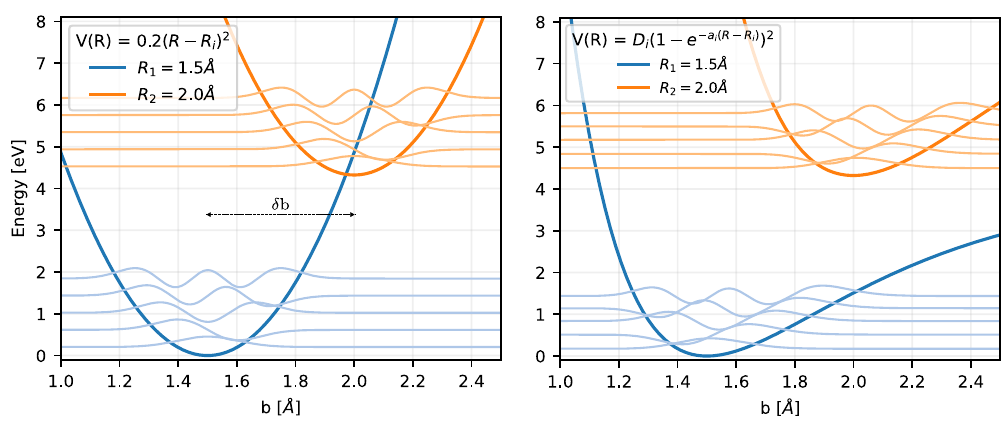} 
\caption{Sketch of the electronic potentials (dark curves) and respective vibrational eigenstates (light curves) shifted by their eigenvalues. a) Harmonic potentials are given by a parabolic equation. b) Morse potentials with dissociation energies $ D_1=4$eV and $ D_2=6$eV, where a$_i = \sqrt{0.15/D_i} $.}
\label{fig:multilevel_schema}
\end{figure}

\begin{figure}[htb]
\centering 
\includegraphics[width=\textwidth]{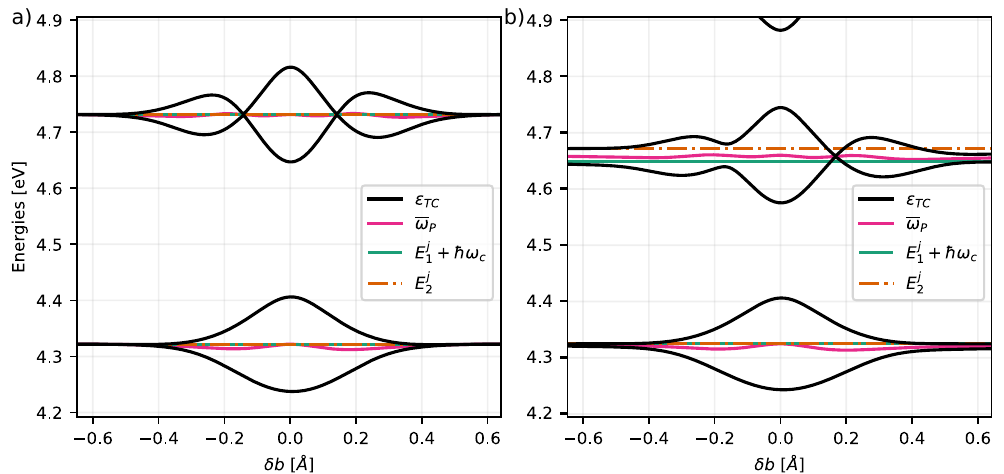} 
\caption{Polaritonic eigenenergies for the multilevel systems with a) equal harmonic potentials and b) different Morse potentials with dissociation energies of $D_1=$~2~eV and $D_2=$~8~eV.}
\label{fig:multilevel_Harm_Morse}
\end{figure}

The observed red shifts in the absorption spectra of the coupled \mgh-cavity system can be explained by the interplay of the inherent anharmonicity of the \glspl{pes} and the geometrical displacement $\delta b$ between the two \glspl{pes}. To investigate how large these two contributions are, we constructed a simplified \gls{tc} model that contains two electronic states, each with 10 vibrational states. To unravel the two effects, we used three different pairs of \glspl{pes} to describe the two electronic states shown in Fig.~\ref{fig:multilevel_schema} and Fig.~\ref{fig:molecular_FC_R}: a) two identical harmonic potentials, b) two Morse potentials with different dissociation energies, and c) the two real molecular \glspl{pes}. In the three cases shown in Fig.~\ref{fig:multilevel_Harm_Morse} and Fig.\ref{fig:molecular_FC_R}, we vary the geometric displacement $\delta b$, determine the transition dipole moments factors (using the molecular transition dipole moment function) for all pairs of vibrational states between the two electronic states, and couple the resulting 20-level system to a single cavity mode. The resulting first four to five polaritonic states as a function of $\delta b$ are plotted in Figs.~\ref{fig:multilevel_Harm_Morse} and ~\ref{fig:molecular_FC_R}. To monitor the asymmetry of the Rabi splitting, we calculate the average energy $\overline{\omega}_P$ of the \gls{lp} and \gls{up} eigenstates. If $\overline{\omega}_P$ is identical to the uncoupled eigenstates (forming \gls{lp} and \gls{up}), the Rabi splitting is perfectly symmetric if its smaller or larger, \gls{lp} and \gls{up} are redshifted or blueshifted, respectively. Even in the case of identical harmonic potentials (Fig.~\ref{fig:multilevel_Harm_Morse}~a)) the Rabi splitting is symmetric only if $\delta b = 0 $ otherwise a weak red shift is observed due to the nonzero contribution from higher vibrational states. The inherent anharmonicity of the Morse potentials only slightly increases the redshift of the Rabi splitting, see Fig.~\ref{fig:multilevel_Harm_Morse}~b).

For the molecular case, we observe a significantly larger redshift of the average polariton frequency for $\delta b \neq 0$ compared to the two model systems, see Fig.~\ref{fig:molecular_FC_R}~a). As the potentials are softer and more asymmetric/anharmonic, the vibrational levels are closer in energy. And as a consequence, the influence of higher states on the \gls{lp} and \gls{up} state is significantly stronger.

\begin{figure}[htb]
\centering 
\includegraphics[width=8.5cm]{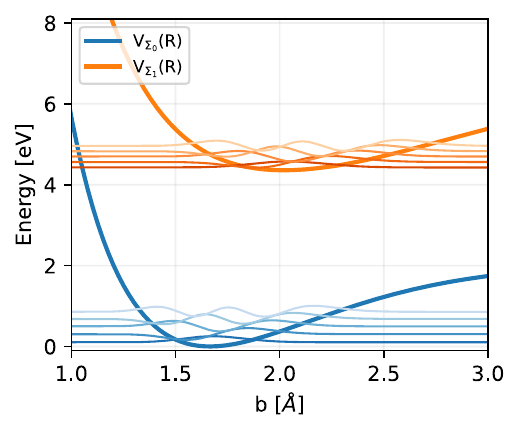} 
\caption{Sketch of the \mgh electronic potentials (dark curves) and respective vibrational eigenstates (light curves) shifted by their eigenvalues.}
\label{fig:multilevel_molecular_pots}
\end{figure}

\begin{figure}[!htb]
	\centering 
	\includegraphics[width=\textwidth]{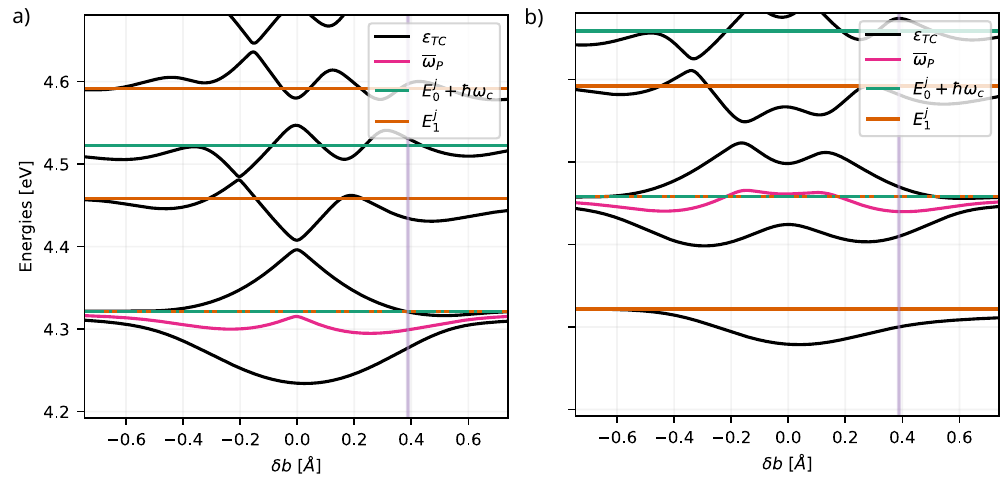}
	\caption{Polaritonic eigenenergies for the multilevel systems with molecular potentials resonant a) with the $\ket{g,v=0}\rightarrow\ket{e,v=0}$ transition and b) at with the $\ket{g,v=0}\rightarrow\ket{e,v=1}$ transition. Purple line denotes the natural relative position between the potentials.}
	\label{fig:molecular_FC_R}
\end{figure}

\begin{figure}[!htb]
	\centering 
	\includegraphics[width=\textwidth]{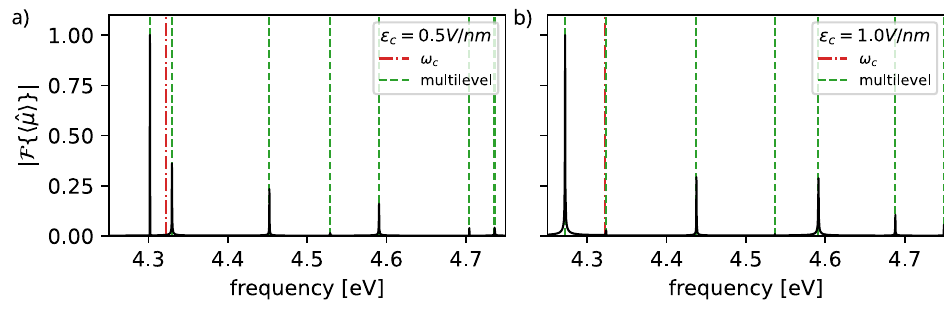}
	\caption{Absorption spectra for the molecular propagations of the \gls{etc} model, compared to the multilevel eigenvalues (green dotted lines). Propagations and multilevel systems considering a cavity frequency of $\omega_c = 4.322$~eV and coupling strengths of a) $\SI{3.5e-3}{au}$ and b) $\SI{6.9e-3}{au}$. }
	\label{fig:mol_abspectrum_multilevel}
\end{figure}

As a benchmark, we compare the absorption spectra peaks for the coupled single molecule system with the eigenvalues of the coupled vibrational multilevel states system in Fig.~\ref{fig:mol_abspectrum_multilevel}, reaching a satisfactory agreement between curves with different vacuum field coupling strengths.

\section{Derivation of the effective molecular Tavis-Cummings model}
\label{sec:TLS}

To reduce computational costs, the molecules coupled to the cavity mode are replaced by effective \glspl{tls}. The two levels represent the first vibrational eigenstate of the two electronic states $\ket{g}$ and $\ket{e}$, separated by $\omega_{eg}$. The individual molecular Hamiltonian reduces to 
\begin{equation}
    H_M =\omega_{eg}\hat\sigma^\dagger\hat\sigma
\end{equation}
The $\vb R$ dependent dipole matrix elements in eq.~\eqref{SI:mu} and the dipole moment squared functions in eq.~\eqref{SI:mu2} are eplaced by the corresponding expectation values:
\begin{align}
\mu_{ij}(\vb R) \rightarrow & \expval{\mu}_{ij} = \bra{\chi_i^0(\vb R)}\mu_{ij} (\vb R)\ket{\chi_j^0(\vb R)}_{\vb R}, \\
\mu_{ij}^2(\vb R) \rightarrow & 
 \expval{\mu^2}_{ij} = \bra{\chi_i^0(\vb R)}\mu^2_{ij}(\vb R)\ket{\chi_j^0(\vb R)}_{\vb R}.
\end{align}
where $\chi_i^0(\vb R)$ is the first vibrational eigenfunction of the electronic state $i$.
In the coherent state basis, the expectation value of the total dipole moment and its quadratic counterpart for a system of $N$-\glspl{tls} have the form
\begin{gather}
\expval{\tilde{\mu}}  = \sum_i^N \expval{\mu}^{(i)} - N\expval{\mu}_{gg},\\
\expval{\tilde{\mu}^2} = \sum_i^N \expval{\mu^2}^{(i)} - 2\expval{\mu}^{(i)} N\expval{\mu}_{gg}  + \sum_{j\neq i}^N  \expval{\mu}^{(j)}\expval{\mu}^{(g)}  + N^2\expval{\mu}_{gg}^2,
\end{gather}
where $i$ and $j$ are the individual contributions from $N$ \glspl{tls}. Since all \glspl{tls} are identical and independent of $\vb R$, the upper index of the dipole and squared dipole moments used to distinguish the individual
molecules are dropped below.

The linear dipole moment couplings within the \gls{rwa} take the following form in the \gls{tls} model:
\begin{align}
\boxthistle{$\bra{G;1} (\hat{a}^\dagger + \hat{a})\expval{\tilde{\mu}} \ket{E^{(a)};0}$} & = \expval{\mu}_{eg} \label{eq:lin1_tls}\\
\boxthistle{$\bra{G;2} (\hat{a}^\dagger + \hat{a}) \expval{\tilde{\mu}} \ket{E^{(a)};1}$} & =\expval{\mu}_{eg} \label{eq:lin2_tls}\\ 
\boxthistle{$\bra{E^{(a)};1} (\hat{a}^\dagger + \hat{a})\expval{\tilde{\mu}} \ket{\mathcal{E}^{(a,b)};0}$} & =\expval{\mu}_{eg} \label{eq:lin3_tls}\\
\boxygreen{$\bra{E^{(a)};1} (\hat{a}^\dagger + \hat{a})\expval{\tilde{\mu}}\ket{E^{(a)};0}$} & = \expval{\mu}_{ee} - \expval{\mu}_{gg} \label{eq:lin4_tls}\\
\end{align}

The diagonal \gls{dse} contribution in the \gls{tls} model reads:
\begin{align}
\bra{G;n} \expval{\tilde{\mu}^2} \ket{G;n}  = & N\expval{\mu^2}_{gg} - N\expval{\mu}^2_{gg}\label{eq:dse1_dia_tls}\\
\bra{E^{(a)};n} \expval{\tilde{\mu}^2} \ket{E^{(a)};n}  = & \left(N-1\right)\expval{\mu^2}_{gg} + \expval{\mu^2}_{ee}  -\left(N-1\right) \expval{\mu}_{ee} \expval{\mu}_{gg} +  \expval{\mu}_{gg}^2 \label{eq:dse2_dia_tls}\\ 
 \bra{\mathcal{E}^{(a,b)};n}  \expval{\tilde{\mu}^2} 
 \ket{\mathcal{E}^{(a,b)};n}  = & \left(N-2\right)  \expval{\mu^2}_{gg} + 2\expval{\mu^2}_{ee}    +2\expval{\mu}_{ee}^2 -2\left(N-2\right)\expval{\mu}_{ee}\expval{\mu}_{gg}+4\expval{\mu}_{gg}^2
    \label{eq:dse3_dia_tls}
\end{align}
The \gls{dse} contributions coupling state which share the same type of \gls{tls} ensemble states read:
\begin{align}
\boxyorange{$ \bra{E^{(a)};n} \expval{\tilde{\mu}^2} \ket{E^{(b)};n}$}  = & \expval{\mu}_{eg}^2\label{eq:dse1_od_s_tls}\\
\boxyorange{$\bra{\mathcal{E}^{(a,b)};n} \expval{\tilde{\mu}^2} \ket{\mathcal{E}^{(b,c)};n}$}  = & \expval{\mu}_{eg}^2\label{eq:dse2_od_s_tls}
\end{align}

\begin{figure}[htb]
	\centering
	\includegraphics[width=8.5cm]{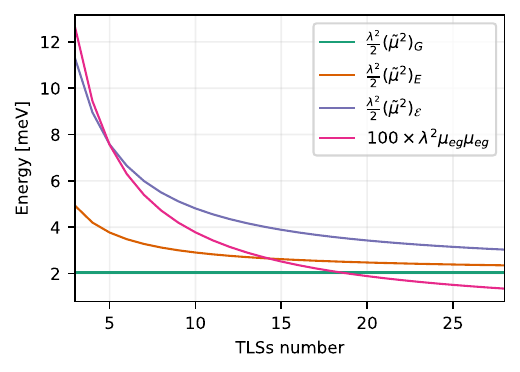}
	\caption{Energy contributions of the diagonal \gls{dse} terms considered as functions of system size $ N $. Coupling strength $\sqrt{N}\lambda_c = \SI{6.9e-3}{au}$ and cavity frequency $\omega_c=4.322$~eV.}
	\label{fig:DSE_matrixelements_N}
\end{figure}

Adding more \glspl{tls} to the system results in more coupling terms in the interaction Hamiltonian. However, due to the scaling of the coupling strength, the energy shift of the diagonal \gls{dse} contributions do not increase with the system size; see Fig.~\ref{fig:DSE_matrixelements_N}. The \gls{dse} ground state shift, given by Eq.~\eqref{eq:dse1_dia_tls}, (blue line in Fig.~\ref{fig:DSE_matrixelements_N}) remains constant, while the other two state-specific shifts, given by Eqs.~\eqref{eq:dse2_dia_tls} and \eqref{eq:dse3_dia_tls} (green and orange lines), decay with one over $N$. However, they converge to a nonzero value. The \gls{dse} contributions coupling state which share the same type of \gls{tls} ensemble states, see Eqs.~\eqref{eq:dse1_od_s_tls} and ~\eqref{eq:dse2_od_s_tls}, are very small and decay to zero with one over $N$.

Due to the simplicity of the \gls{tls}, we have analytical formulas to describe the  population dynamics of its states and absorption spectrum. The Rabi frequency for the first manifold excitation states, which describes the oscillations between ground and excited states populations, is given by
\begin{equation} \label{eq:OmegaR}
    \Omega_R = \sqrt{2N\omega_c(\lambda_c\mu_{eg})^2+\qty(\Delta\omega_c)^2},
\end{equation}
where $\Delta\omega_c=\omega_{eg}-\omega_c$ is the cavity detuning to the \gls{tls} excitation frequency and $N$ is the number of emitters. The polariton states eigenenergies are given by
\begin{equation} \label{eq:Epolariton}
    E_{\pm} = \frac{(\omega_c+\omega_{eg})}{2} \pm \frac{\Omega_R}{2},
\end{equation}
where $\ket{UP}\equiv\ket{+}$ and $\ket{LP}\equiv\ket{-}$.

\begin{figure}[htb] 
\centering
\includegraphics[width=\textwidth]{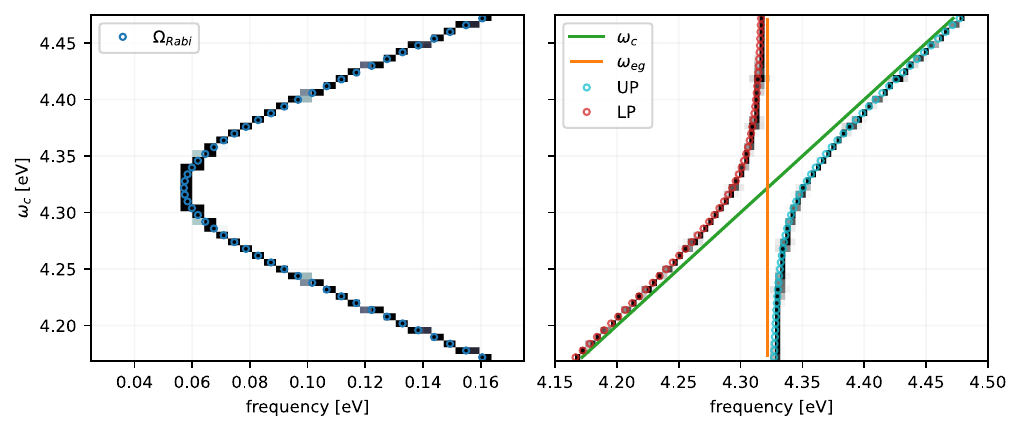}
\caption{Absorption spectrum for a $2$-TLS system with respect to cavity frequency $\omega_c$ using a $\omega_{eg}^{TLS}=4.322$eV and a coupling strength of $\sqrt{N}\lambda_c = \SI{6.9e-3}{au}$. Left panel shows the Rabi frequencies compared to Eq.~\eqref{eq:OmegaR}, and right panel shows the electronic transitions for the \gls{up} and \gls{lp} states, compared to Eq.~\eqref{eq:Epolariton}.}
\label{fig:FT_TLS}
\end{figure}

The absorption spectra for a \gls{tls} can be calculated analogously to the molecular system case, and an example dispersion curve for the \gls{tls} model is shown in Fig.~\ref{fig:FT_TLS}.
The polariton state peaks follow hyperbolic curves whose asymptotes are the \gls{tls} excitation frequency $\omega_{eg}^{TLS}$ and the cavity frequency $\omega_c^{TLS}$, centered at the resonance point.

\begin{figure}[htb]
\centering
\includegraphics[width=\textwidth]{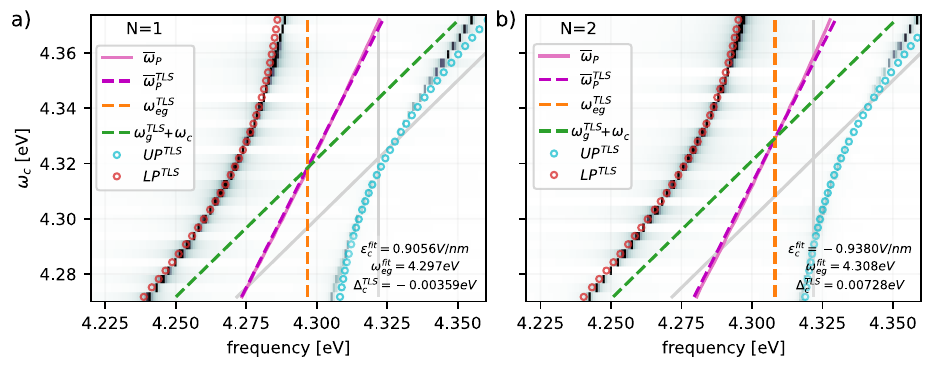}
\caption{Fitting of \glspl{tls} upper and lower polariton curves using the molecular dispersion curve for the case of a) one and b) two \mgh molecules in the region where the middle polariton frequency $\overline\omega_P$ is almost linear. The fitted \gls{tls} polariton curves have asymptotes that correspond to the optimized emitter resonance energy $\omega_{eg}^{TLS}$, and to the ground level energy $\omega_g^{TLS}$ plus cavity excitation which gives the detuning $\Delta_c^{TLS}$. The gray lines correspond to the molecular parameters of $\omega_{eg}$ and $\omega_g+\omega_c$. }
\label{fig:TLS_optimization}
\end{figure}

By definition, the \glspl{tls} model is characterized by a symmetric Rabi splitting, see Fig.~\ref{fig:FT_TLS}, and cannot reproduce the observed asymmetry in the polaritonic eigenstates; see Section~\ref{sec:spectra}. Based on the eigenvalues of the coupled molecule-cavity system, it is possible to optimize the \gls{tls} to mimic the energetics of the real molecular polariton states. Starting from the dispersion curve of the molecule-cavity system, see Fig.~\ref{fig:TLS_optimization} for the cases of one and two \mgh, we define the region where $\overline{\omega}_P$ (pink line) is approximately linear. In the next step, we determine the energy gap between the two levels $\omega_{eg}^{TLS}$ and an effective detuning of the cavity frequency $\Delta_c^{TLS}$  with respect to the one used for the molecular setup. By defining a suitable $\overline{\omega}_P^{TLS}$ (pink dashed line Fig.~\ref{fig:TLS_optimization}), we can compute the calculated $\omega_{eg}^{TLS}$ and $\omega_g^{TLS}+\omega_c$ as the vertical asymptote (orange dashed line) and the diagonal asymptote (green dashed line), respectively. Furthermore, we also defined an optimal coupling strength $\lambda_c^{TLS}$ from the Rabi frequency peak in the absorption spectra to replicate the molecular results. 

All optimized \gls{tls} parameters for one, two, and three \mgh coupled to a cavity are listed in table \ref{tab:opt_pars}. We use the optimized parameter of the three-molecule case as default for larger ensembles of \glspl{tls} considering that the resonance frequency seems to converge in the molecular case as the system grows.

\begin{table}[!htb]    
	\centering
	\caption{Optimization parameters for the \glspl{tls} fitted in the molecular absorption spectra used in the main text.}
	\label{tab:opt_pars}
	\begin{tabular}{l|rrr}
		N & 1 & 2 & 3  \\ \hline
            $\lambda_c^{TLS}$ [$10^{-3}$au] &  6.25\quad & 6.47\quad & 6.52 \\ 
		$\omega_{eg}^{TLS}$ [eV] &  4.297 & 4.308 & 4.312 \\ 
		$\Delta_c^{TLS}$ [meV] & -3.590 & 7.280 & 11.60  
	\end{tabular}
\end{table}

The curves of Fig.~10 and Fig.~11(a) were fitted to the formula $f(N) = a + bN^c$ using the nonlinear least squares method. The covariance matrices for the fitted curves of Max.~$\Delta P$, $\Delta E_\text{LP}$, $\Delta E_\text{D}$ and $\Delta E_\text{UP}$ are given in the following equations, respectively, where the square root of the diagonals gives the standard deviation errors on the parameters $a$, $b$ and $c$.

\begin{equation} \label{covariance_DP}
\begin{pmatrix} 
            \SI{2.04261431e-10}{} & \SI{1.44747431e-09}{} & \SI{-1.07616473e-08}{}  \\ 
            \SI{1.44747431e-09}{} & \SI{ 1.52196463e-08}{} & \SI{-9.29061251e-08}{} \\ 
		\SI{-1.07616473e-08}{} & \SI{ -9.29061251e-08}{} &\SI{6.30274653e-07}{} 
\end{pmatrix} ,
\end{equation}
\begin{equation} \label{covariance_DELP}
\begin{pmatrix} 
            \SI{1.16410899e-15}{} & \SI{ 8.99753262e-15}{} & \SI{ -6.22680416e-12}{}  \\ 
            \SI{8.99753262e-15}{} & \SI{ 1.03397203e-13}{} & \SI{-5.89935765e-11}{} \\ 
		\SI{-6.22680416e-12}{} & \SI{-5.89935765e-11}{} &\SI{3.72584023e-08}{}
\end{pmatrix},
\end{equation}
\begin{equation} \label{covariance_DED}
\begin{pmatrix} 
            \SI{1.16410899e-15}{} & \SI{8.99753262e-15}{} & \SI{-6.22680416e-12}{}  \\ 
            \SI{8.99753262e-15}{} & \SI{1.03397203e-13}{} & \SI{ -5.89935765e-11}{} \\ 
		\SI{-6.22680416e-12}{} & \SI{-5.89935765e-11}{} &\SI{3.72584023e-08}{}
\end{pmatrix},
\end{equation}
\begin{equation} \label{covariance_DEUP}
\begin{pmatrix} 
            \SI{1.16410899e-15}{} & \SI{ 8.99753262e-15}{} & \SI{-6.22680416e-12}{}  \\ 
            \SI{8.99753262e-15}{} & \SI{ 1.03397203e-13}{} & \SI{-5.89935765e-11}{} \\ 
		\SI{-6.22680416e-12}{} & \SI{ -5.89935765e-11}{} &\SI{ 3.72584023e-08}{}
\end{pmatrix} .
\end{equation}

\clearpage

\section{Resolution of identity approach for the squared dipole operators}

\begin{figure}[htb]
    \centering
    \includegraphics[width=.95\textwidth]{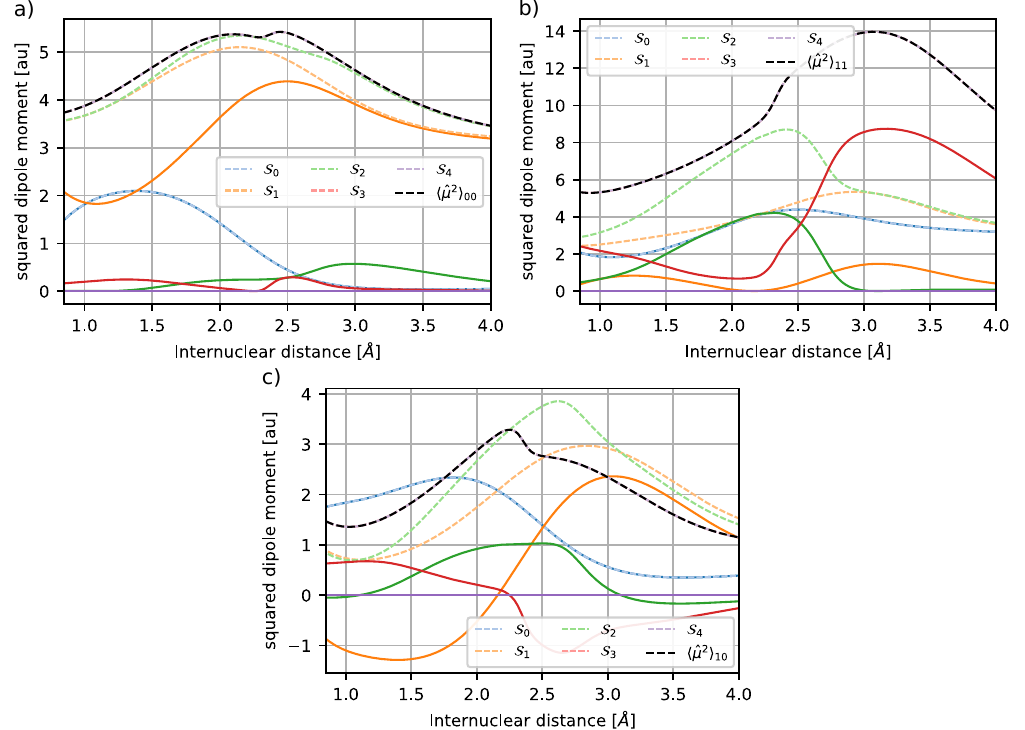}
    \caption{Resolution of identity for the squared dipole moments of a) $\mu^2_{gg}=\ev{\hat\mu^2}_{00}$, b) $\mu^2_{ee}=\ev{\hat\mu^2}_{11}$ and c) $\mu^2_{eg}=\ev{\hat\mu^2}_{10}$, where $\mathcal{S}_m$ is given by $\Sigma_i^m \mu_{0i}\mu_{i0}$, $\Sigma_i^m \mu_{1i}\mu_{i1}$ and $\Sigma_i^m \mu_{1i}\mu_{i0}$, respectively, and darker curves correspond to the final terms of the partial summations.}
    \label{fig:mu2_convergence}
\end{figure}

The squared dipole operators used in the \gls{dse} coupling terms were calculated by the resolution of the identity approach given in Eq.~18. The convergence of the summation is shown in Fig.~\ref{fig:mu2_convergence}, where $\mathcal{S}_m$ refers to the partial summation on the electronic states of $\ev{\hat\mu^2}_{kl}=\sum^m_{i=0}\mu_{ki}\mu_{il}$. As can be seen from the plots, the convergence of the summations are satisfactory using only the first five electronic states. 

\bibliography{lit.bib}